\documentclass[aps,prl,preprint,superscriptaddress]{revtex4-2}

\usepackage[colorlinks=true,linkcolor=blue]{hyperref}
\usepackage{graphicx} 
\DeclareGraphicsExtensions{.pdf,.png,.jpg}

\usepackage{ragged2e}
\usepackage[font=small,labelfont=bf,justification=RaggedRight]{caption}
\DeclareCaptionLabelSeparator{bar}{\space\textbar\space}
\usepackage{amsmath}
\usepackage{siunitx}
\usepackage{bm}
\usepackage{amsfonts}
\usepackage{amssymb}
\usepackage[utf8]{inputenc}
\usepackage[T1]{fontenc}
\usepackage{mathptmx}
\usepackage{etoolbox}
\usepackage{mathtools}
\usepackage[subtle]{savetrees}
\usepackage{setspace}

\usepackage{comment}
\excludecomment{outlinesec}

\setcitestyle{super}

\usepackage[normalem]{ulem} 
\usepackage{color}

\renewcommand{\figurename}{\textbf{Fig.}}
\renewcommand{\thefigure}{{\bf \arabic{figure}}}

\DeclareSIUnit\angstrom{\text {Å}}
\DeclareSIUnit\ions{\text {ions}}

\usepackage{lineno}
\usepackage{orcidlink}

\begin{document}

\title{Long-Range Structural and Magnetic Coherence \\ in Embedded Mesospin Metamaterials}

\author{Christina Vantaraki \orcidlink{0000-0001-5772-4649}}
\affiliation{Department of Physics and Astronomy, Uppsala University, Box 516, 75120 Uppsala, Sweden}

\author{Oier Bikondoa \orcidlink{0000-0001-9004-9032}}
\affiliation{Department of Physics, University of Warwick, Coventry CV4 7AL, United Kingdom}

\author{Matías P. Grassi \orcidlink{0000-0001-9551-9793}}
\affiliation{Department of Physics and Astronomy, Uppsala University, Box 516, 75120 Uppsala, Sweden}

\author{Brindaban Ojha \orcidlink{0000-0002-0276-6937}}
\affiliation{Department of Physics and Astronomy, Uppsala University, Box 516, 75120 Uppsala, Sweden}

\author{Alkaios Stamatelatos \orcidlink{0000-0001-8658-8316}}
\affiliation{Department of Physics, University of Warwick, Coventry CV4 7AL, United Kingdom}

\author{Natalia Kwiatek-Maroszek \orcidlink{0000-0001-6861-4066}}
\affiliation{ALBA Synchrotron Light Facility, 08290-Cerdanyola del Valles, Barcelona, Spain}

\author{Miguel Angel Niño Orti \orcidlink{0000-0003-3692-147X}}
\affiliation{Instituto de Química Física "Blas Cabrera", CSIC, 28006 Madrid, Spain}

\author{Michael Foerster \orcidlink{0000-0002-4147-6668}}
\affiliation{ALBA Synchrotron Light Facility, 08290-Cerdanyola del Valles, Barcelona, Spain}

\author{Thomas Saerbeck \orcidlink{0000-0001-7913-691X}}
\affiliation{Institut Laue-Langevin, Grenoble, France}

\author{Daniel Primetzhofer \orcidlink{0000-0002-5815-3742}}
\affiliation{Department of Physics and Astronomy, Uppsala University, Box 516, 75120 Uppsala, Sweden}

\author{Max Wolff \orcidlink{0000-0002-7517-8204}}
\affiliation{Department of Physics and Astronomy, Uppsala University, Box 516, 75120 Uppsala, Sweden}

\author{Nicolas Jaouen \orcidlink{0000-0003-1781-7669}}
\affiliation{Synchrotron SOLEIL, L’Orme des Merisiers
Saint-Aubin, Gif-sur-Yvette 91192, France}

\author{Thomas P. A. Hase \orcidlink{0000-0001-5274-5942}}
\affiliation{Department of Physics, University of Warwick, Coventry CV4 7AL, United Kingdom}

\author{Vassilios Kapaklis \orcidlink{0000-0002-6105-1659}}
\affiliation{Department of Physics and Astronomy, Uppsala University, Box 516, 75120 Uppsala, Sweden}

\maketitle

\noindent
\footnotesize
Corresponding author: Vassilios Kapaklis (\href{vassilios.kapaklis@physics.uu.se}{vassilios.kapaklis@physics.uu.se}).
\normalsize

    \section{Abstract}

{\bf
Engineered assemblies of interacting magnetic elements—magnetic metamaterials—provide a powerful route to tailor collective magnetic order and dynamics \cite{Heyderman:2013gb, Nisoli_rev_2013}. By structuring matter at the mesoscale, they bridge atomic magnetism and macroscopic functionality, enabling emergent behaviour inaccessible in conventional materials\cite{Nisoli:2017hg, Rougemaille:2019ef, Skjearvo_2020}. However, realizing large-area metamaterials that combine high morphological uniformity with intrinsic long-range order has remained challenging, largely due to the structural disorder inherent to lithographic fabrication. Here we demonstrate a scalable route to structurally and magnetically coherent metamaterials by embedding iron-ions to form mesospins within a non-magnetic thin film palladium host matrix. Using controlled implantation, we realize morphologically uniform arrays that spontaneously develop extended antiferromagnetic order in the as-fabricated state — without the need of external annealing or field cycling. Resonant X-ray scattering and microscopy reveal sharp magnetic Bragg peaks modulated by the mesospin form factor, evidencing long-range antiferromagnetic order coupled to structural coherence. This embedded architecture establishes a platform for exploring coherent spin–photon interactions and functional X-ray scattering in magnetic metamaterials free from lithographic topography and disorder.
}

    \section{Introduction}

Magnetic metamaterials have emerged as a versatile platform for exploring how geometry, dimensionality, and designed interactions govern collective magnetic behavior \cite{Heyderman:2013gb, Rougemaille:2019ef}. Through tailored arrangements of interacting mesoscopic elements, systems can be designed to incorporate frustrated \cite{SchifferP2006Aii} and correlated magnetic states \cite{Nisoli:2017hg} that bridge atomic-scale magnetism and macroscopic functionality. This capacity to engineer interactions at the mesoscale has enabled advances in thermally driven ordering \cite{KapaklisVassilios2012Masi, Zhang:2013ga, KapaklisVassilios2014Tfia}, magnetic phases and kinetics defined by topology \cite{Schiffer_Science_2023, Schiffer_PNAS_2025}, reconfigurable spin textures \cite{Skovdal_collapse_2021, Nanny_textures_2022}, magnonic circuitry \cite{Gartside_RC, Kaffash_Review_2021}, reconfigurable logic \cite{Gypens:2018br, Arava:2019kq, astroid_clocking} and hybrid magneto-optical systems \cite{OAM_ASI_PRL, McCarterMargaretR.2023Arcp}, where geometry and coupling strength can be tuned independently of intrinsic material parameters. Despite this progress, fabricating large-area arrays that combine morphological uniformity with intrinsic long-range magnetic order remains challenging. Conventional lithographic routes introduce edge roughness, inter-element variability, and chemical inhomogeneity that disrupt long-range correlations and obscure intrinsic magnetic order. These imperfections often require post-growth annealing or field cycling to reveal ordered states, potentially masking the spontaneous emergence of collective order. Conventional lithographically defined systems are also constrained by the intrinsic properties of the base materials from which the metamaterials are fabricated. Key parameters—such as the magnetic ordering temperature, anisotropy, or moment magnitude—can typically only be modified indirectly, for instance through finite-size effects or careful control during film deposition. As a result, the available design space is constrained, and the functional versatility of the resulting metamaterials limited.


\begin{figure}[t]
\includegraphics[width=0.55\linewidth]{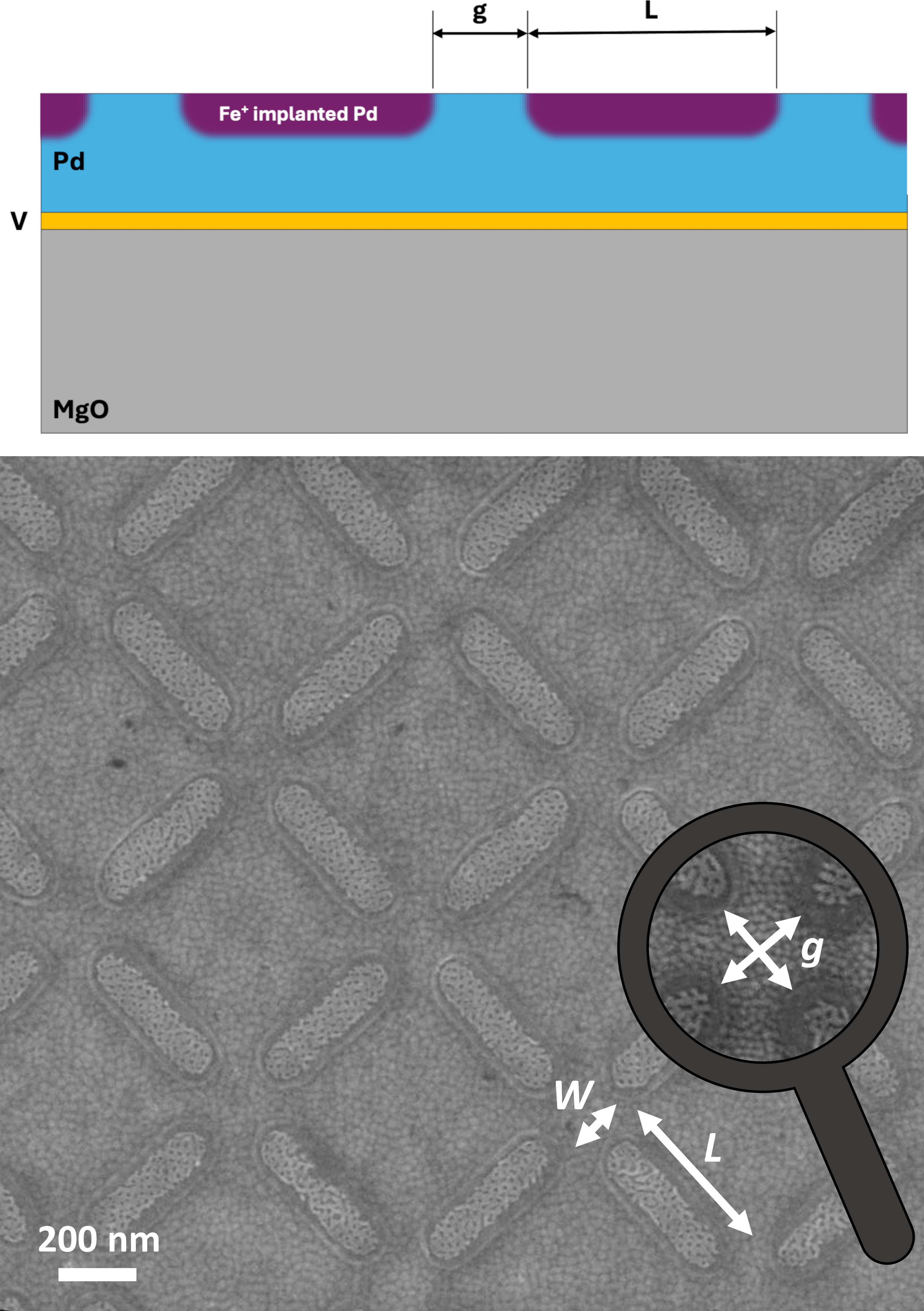}
\caption{\linespread{1.0} \footnotesize {\bf Implanted magnetic metamaterial.} Top: Schematic cross-section of the implanted metamaterial architecture \cite{VantarakiChristina2024Mmbi}. Bottom: Scanning electron microscopy image of a square artificial spin ice patterned via ion implantation, illustrating the high uniformity of ion-implanted mesospins. The surface morphology of the array reflects the patterned implantation mask, and a slight surface swelling of the implanted regions is visible due to the local lattice expansion induced by Fe incorporation. The mesospins of Fe-implanted regions in a paramagnetic Pd matrix, have lateral dimensions of $L=$\SI{470}{nm}, $W=$\SI{170}{nm} and an edge-to-edge gap of $g=$\SI{170}{nm}. Details of the sample fabrication are presented in Methods.}
\label{Figure1}
\end{figure}

Advances in resonant and coherent X-ray scattering now offer a powerful route to probe magnetic metamaterials with high sensitivity to both chemical and magnetic correlations. These techniques enable direct access to spin textures, interference effects, and topological configurations that remain otherwise hidden in standard dichroic approaches \cite{OAM_ASI_PRL, McCarterMargaretR.2023Arcp}. Fully exploiting this capability, however, requires metamaterials with well-defined morphology and robust long-range magnetic order—conditions that enhance coherence, sharpen reciprocal-space features, and expose the underlying structure–factor physics. This has intensified the need for material architectures that are intrinsically compatible with coherent scattering approaches and capable of supporting reproducible emergent magnetic order across extended areas.


Here, we present a scalable approach to fabricating coherent magnetic metamaterials by embedding ferromagnetic mesospins within a non-magnetic host matrix. For demonstration, we employ a square artificial spin ice (ASI) geometry --- a popular lattice with a well-defined antiferromagnetic ground state (Fig. \ref{Figure1}) \cite{KapaklisVassilios2014Tfia, Skjearvo_2020, Nisoli:2017hg}. Controlled Fe$^+$ ion implantation into Pd thin films produces single-domain mesospins that spontaneously assemble into extended antiferromagnetic domains in the {\it as-implanted} state. Relative to our earlier implantation studies that established fabrication and real-space magnetic textures \cite{VantarakiChristina2024Mmbi, Vantaraki_inhomogeneities}, the present work provides a reciprocal-space demonstration of simultaneous structural and magnetic long-range coherence. Resonant X-ray reflectometry, polarized neutron reflectometry, and photoemission electron microscopy establish the structural and magnetic depth profiles, while resonant soft X-ray diffraction reveals sharp structural Bragg peaks together with mixed-parity magnetic reflections and a clearly resolved mesospin-basis (``$\times$'') envelope. Together, these findings establish a class of structurally embedded magnetic metamaterials with intrinsic long-range order, providing a basis for quantitatively interpretable, geometry- and form-factor-encoded scattering studies of collective magnetism and coherent photon--spin coupling.

\section{\label{sec:results} Results and discussion} 

\subsection{\label{subsec:reflectivity} Structural and magnetization depth profiles} 

\begin{figure}[t]
\includegraphics[width=0.55\linewidth]{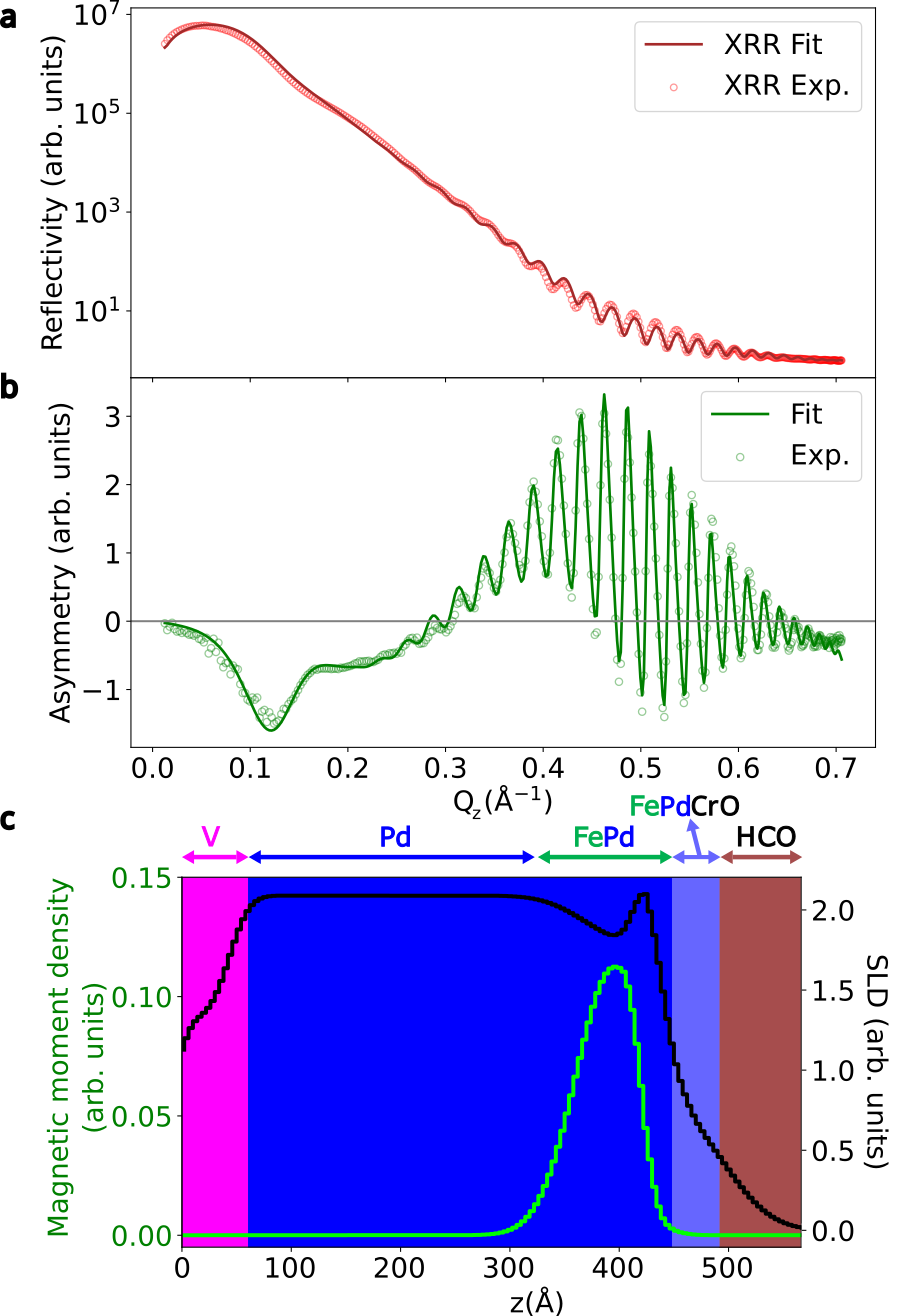}
\caption{\linespread{1.0} \footnotesize {\bf Resonant X-ray reflectivity and magnetic moment profiles.} {\bf a} X-ray reflectivity curve. {\bf b} Asymmetry obtained from the subtraction of XRR signals acquired with opposite in-plane magnetic fields with a beam energy of \SI{707}{eV}, sensitive to Fe. {\bf c} Electron density (black line) and magnetic moment density (green line) extracted from simultaneously fitting the reflectivity and asymmetry data for different X-ray photon energies. The depth profiles originate at the MgO substrate ($z = 0$), followed by the V adhesion layer, Pd film, and Fe-implanted region. These layer positions are indicated in the panel for clarity.}
\label{FigureXrays}
\end{figure}
To establish the structural and magnetization depth profiles of the implanted films, we employed resonant X-ray reflectivity (XRR) and polarized neutron reflectivity (PNR) on continuous-film samples. Palladium lies close to the Stoner criterion for ferromagnetism and can become magnetically polarized when surrounded by magnetic impurities \cite{Vogel:1997bw, HaseThomasP.A.2014Peod, PapaioannouEvangelosTh2010Dace}. Fig.~\ref{FigureXrays}a and \ref{FigureXrays}b show the total reflectivity and asymmetry signals for left ($I^{L}$) and right ($I^{R}$) circularly polarized X-rays, measured at the Fe $L_{3}$ edge (\SI{707}{eV}) from the sample with a nominal \SI{40}{nm} Pd film. A simultaneous fit of these data to a single model yields both the electron density (scattering-length density, SLD) and the Fe magnetic moment density profiles, presented in Fig.~\ref{FigureXrays}c. These profiles reproduce the expected layer sequence: MgO substrate, V adhesion layer, Pd film, Fe-implanted region, a thin PdCrO layer, and a surface water/hydrocarbon contamination layer—demonstrating that the implantation process preserves the overall film morphology. The apparent thickness of the contamination layer in Fig.~\ref{FigureXrays}c may be somewhat overestimated, as the fitting model can also capture contributions from top-layer roughness induced by ion bombardment during the ion implantation process (see Fig.~\ref{Figure1}). The Fe concentration peaks near the surface and extends approximately \SI{15}{nm} into the Pd layer. The magnetic moment profile shows a slightly asymmetric but well-defined distribution, peaking roughly \SI{5}{nm} below the surface with a width of about \SI{7}{nm}. A notable feature is the dip in SLD, and therefore in electron density, within the mixed Pd:Fe region (Fig.~\ref{FigureXrays}c). In resonant reflectivity this effective SLD is energy dependent and includes dispersive corrections ($f'$, $f''$) near the Fe $L_{3}$ edge; accordingly, the dip reflects both the implantation-driven composition/density change and the resonant modification of the complex atomic scattering factors at the measurement energy. This subtle difference in electron density between the Pd matrix and the Fe-implanted regions provides the contrast used to explore the implanted metamaterials.

Polarised neutron reflectometry analyses (Supplementary Information) reproduce similar magnetic profiles, but in this case the magnetic region not only coincides with the implanted Fe distribution but also includes contributions from induced Pd polarization, consistent with previous ion-beam studies\cite{StromPetter2022Soft, VantarakiChristina2024Mmbi, Vantaraki_inhomogeneities}. The two probes are therefore complementary: resonant X-ray methods at the Fe edge are Fe-selective, whereas PNR reports the total magnetic depth profile (Fe+Pd). For these implantation-defined structures, resonant XRR/PNR comparison is most physically meaningful at the level of the extracted chemical and magnetic SLD depth profiles, rather than by tabulating single thickness/roughness values that act only as model-dependent effective parameters for graded interfaces. A key finding, when considering both the X-ray and neutron data (Supplementary Information), is the consistency between two independently prepared samples, probed using distinctly different techniques, which demonstrates the reproducibility of the ion implantation process in generating well-defined magnetic volumes close to the surface.

\subsection{\label{subsec:microscopy} Magnetic microscopy}

\begin{figure}[t]
\includegraphics[width=0.4\linewidth]{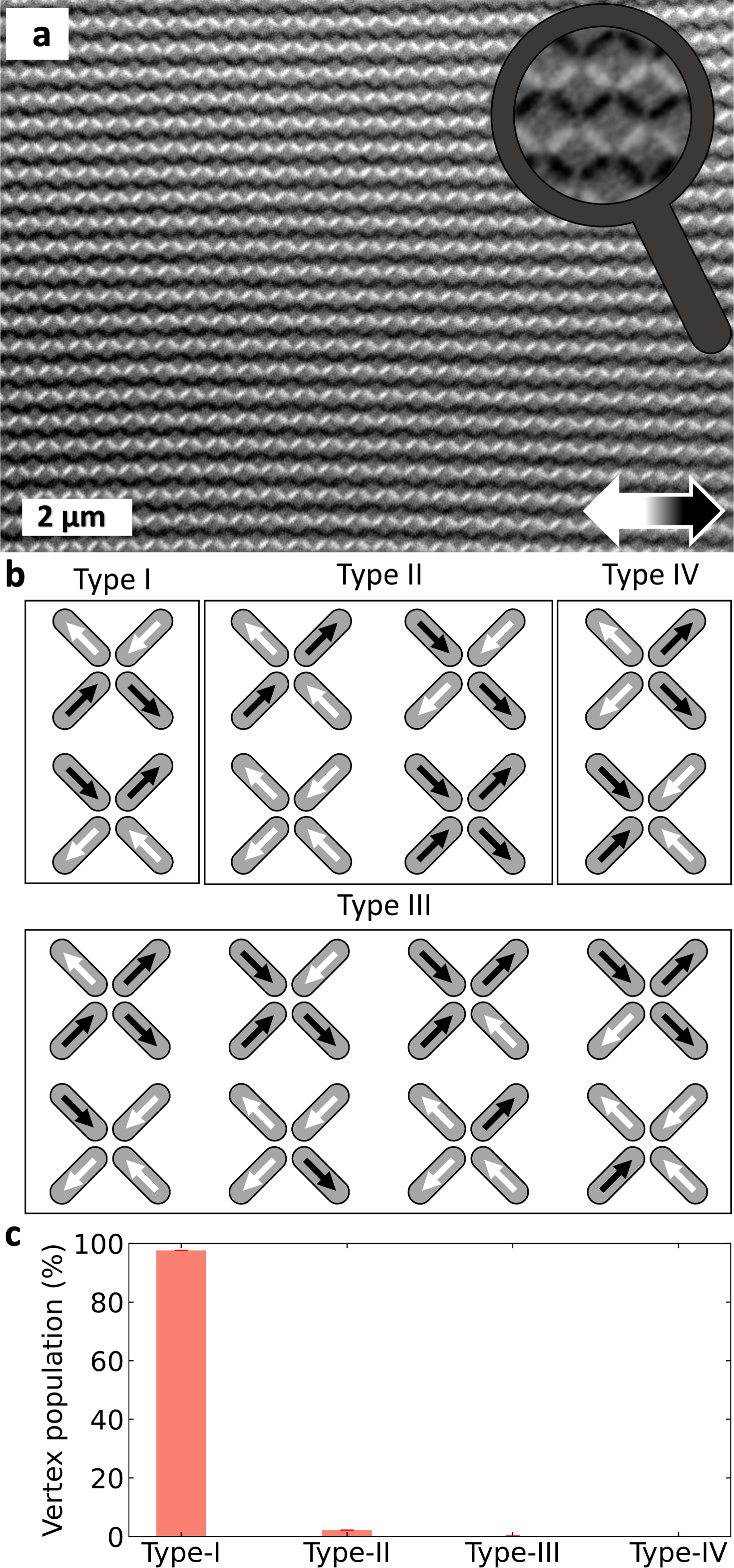}
\caption{\linespread{1.0} \footnotesize {\bf Real-space magnetic imaging and vertex populations.} {\bf a} Representative PEEM - XMCD image of the implanted square ASI lattice in the as-implanted state. The black and white colors indicate a magnetization component parallel and antiparallel
to the X-ray beam, respectively. {\bf b} Vertex types for the square ASI lattice. Type-I has the lowest energy state, followed by Type-II, III and IV. {\bf c} Vertex population for the implanted square ASI lattice. A dominant Type-I configuration of the ASI lattice is observed, consistent with the expected nearest-neighbour square-ASI ground state and with the direct PEEM--XMCD real-space evidence of extended ordered domains, enabling the use of these samples as benchmarks in X-ray scattering studies.}
\label{PEEM-XMCD}
\end{figure}

\begin{figure*}[t]
\includegraphics[width=\textwidth]{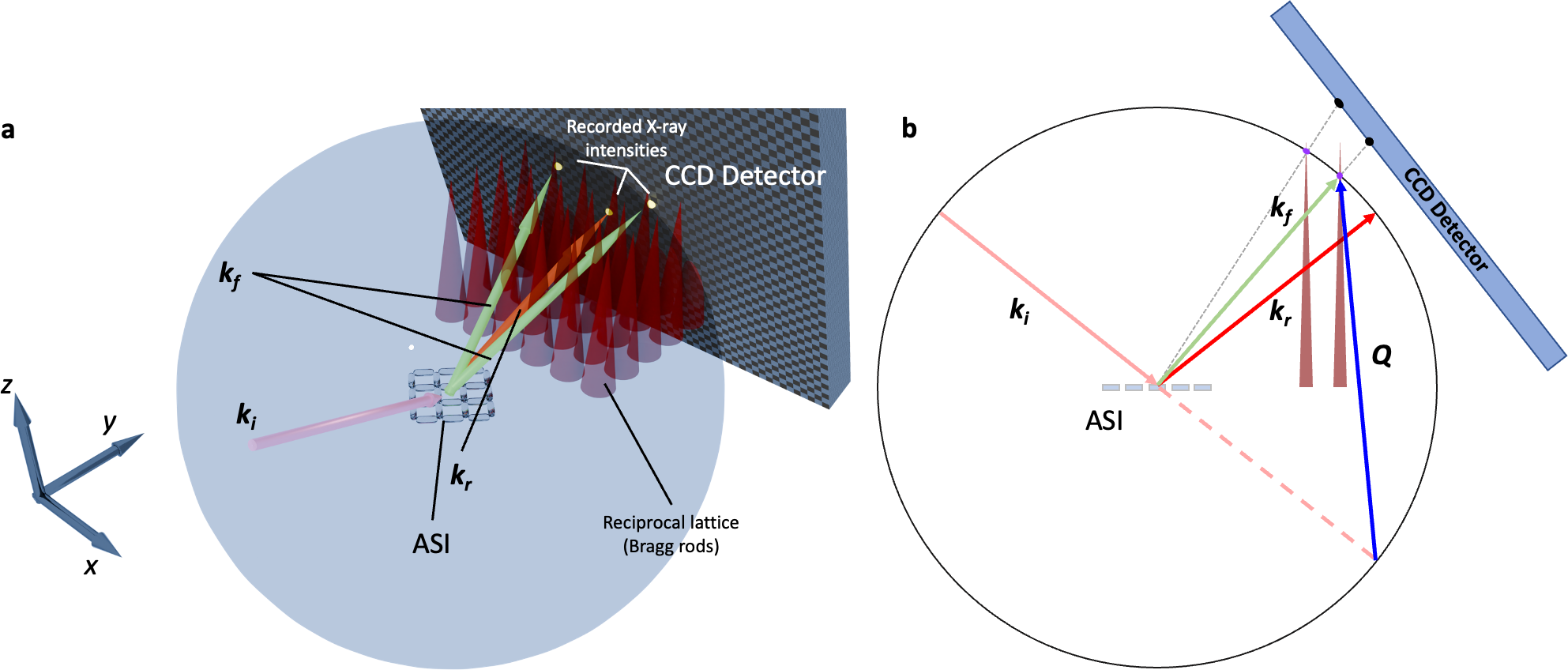}
\caption{\linespread{1.0} \footnotesize {\bf X-ray scattering from implanted ASIs.} {\bf a} Ewald construction illustrating the scattering geometry in the soft X-ray diffraction experiments on the implanted artificial spin ice lattices. The reciprocal lattice of the ASI gives rise to a series of  intensity rods (schematically depicted as cones) extending along the out-of-plane direction. The intensity modulation along these rods encodes information about the vertical structure of the mesospins. The incident beam vector $\mathbf{k}_i$, reflected beam $\mathbf{k}_r$, and scattered beam $\mathbf{k}_f$ define the scattering geometry, with the end of $\mathbf{k}_f$ lying on the surface of the Ewald sphere. A constructive scattering condition is satisfied whenever the Ewald sphere intersects an intensity rod. The recorded diffraction intensities are determined not only by the structural contribution but also by the detector position (gray box), the form factor of the implanted mesospins, and the specific magnetization textures within the elements. {\bf b} Side view ($yz$ plane) of the Ewald construction, showing the intersection of the Ewald sphere with the Bragg rods and the resulting mapping of diffraction peaks onto the detector plane.}
\label{Ewald}
\end{figure*}

Having established the magnetic depth profile of the continuous films, we examine the in-plane magnetic ordering in the implanted array. Photoemission electron microscopy with X-ray magnetic circular dichroism (PEEM–XMCD) provides direct, element-specific imaging of the magnetization at the mesoscale (Fig.~\ref{PEEM-XMCD}a). The PEEM image shows that each implanted element exhibits uniform contrast, confirming single-domain behavior. Moreover, the elements form extended antiferromagnetic domains corresponding to the Type-I ground state of the square geometry (Fig.~\ref{PEEM-XMCD}b). The observation of large, defect-free domains indicates strong inter-element coupling and suggests that the array effectively thermalizes during Fe$^+$ ion dose accumulation with transient diffusion processes allowing the magnetic moments to relax into low-energy configurations before freezing. Such extended, low-defect Type-I domain formation is typically observed only when inter-island interactions are sufficiently strong to overcome local disorder and thermal/field-history effects. This spontaneous ordering mechanism yields a degree of magnetic coherence rarely achieved without post-growth treatment in conventionally patterned systems \cite{MarrowsChristopherH2011Tgoa, Zhang:2013ga, KapaklisVassilios2014Tfia, Arnalds_2D_Ising}.

Statistical analysis of the vertex configurations extracted from the PEEM images reveals an overwhelmingly dominant population of Type-I vertices (Fig.~\ref{PEEM-XMCD}c), confirming that the system naturally adopts its ground-state tiling. The prevalence of low-energy configurations in the {\it as-prepared} arrays, exceeds that of conventionally fabricated artificial spin ices, reflecting the enhanced uniformity and coupling strength achieved through the embedded architecture. The effective thermalization achieved during ion implantation yields stronger magnetic ordering than typically observed in conventionally fabricated arrays made from $\delta$-doped Pd(Fe) \cite{KapaklisVassilios2014Tfia,OstmanErik2018Imia} or permalloy \cite{SchifferP2006Aii}, and it is comparable to the behaviour reported in FePd$_{3}$-based ASIs \cite{DriskoJasper2015Fsaa} after post-fabrication treatment.

\subsection{\label{subsec:XRMS} Resonant X-ray scattering}

Diffraction is an alternative probe to microscopy for studying the arrays, yielding direct information on correlations. We start by considering nonresonant scattering using an X-ray energy of \SI{690}{eV}. At this energy, the measurement is sensitive only to charge (electron density) correlations. The difference between the incident $\mathbf{k}_i$ and scattered $\mathbf{k}_f$ wave vectors, $\mathbf{Q} = \mathbf{k}_f - \mathbf{k}_i$, defines the momentum transfer. The scattering vector, $\mathbf{Q}$, can be decomposed into the orthogonal laboratory-frame components $Q_x$, $Q_y$, and $Q_z$ defined using the diffractometer axes. The $Q_x$ and $Q_y$ components correspond to the in-plane directions (with $Q_y$ aligned with the beam direction), whereas $Q_z$ represents the out-of-plane direction with respect to the sample surface. The specific dimensions of the mesospin islands directly determine the scattering intensity distribution in reciprocal space. The well-defined lateral periodicities of the array, give rise to sharp peaks in  $Q_x$ and $Q_y$ but since the vertical extent of the islands is finite and not periodic, the scattered intensity manifests as rods of scattering as a function of $Q_z$ with an intensity profile resembling that of the reflectivity shown in Fig.~\ref{FigureXrays}.

\begin{figure}[t]
\includegraphics[width=0.8\textwidth] {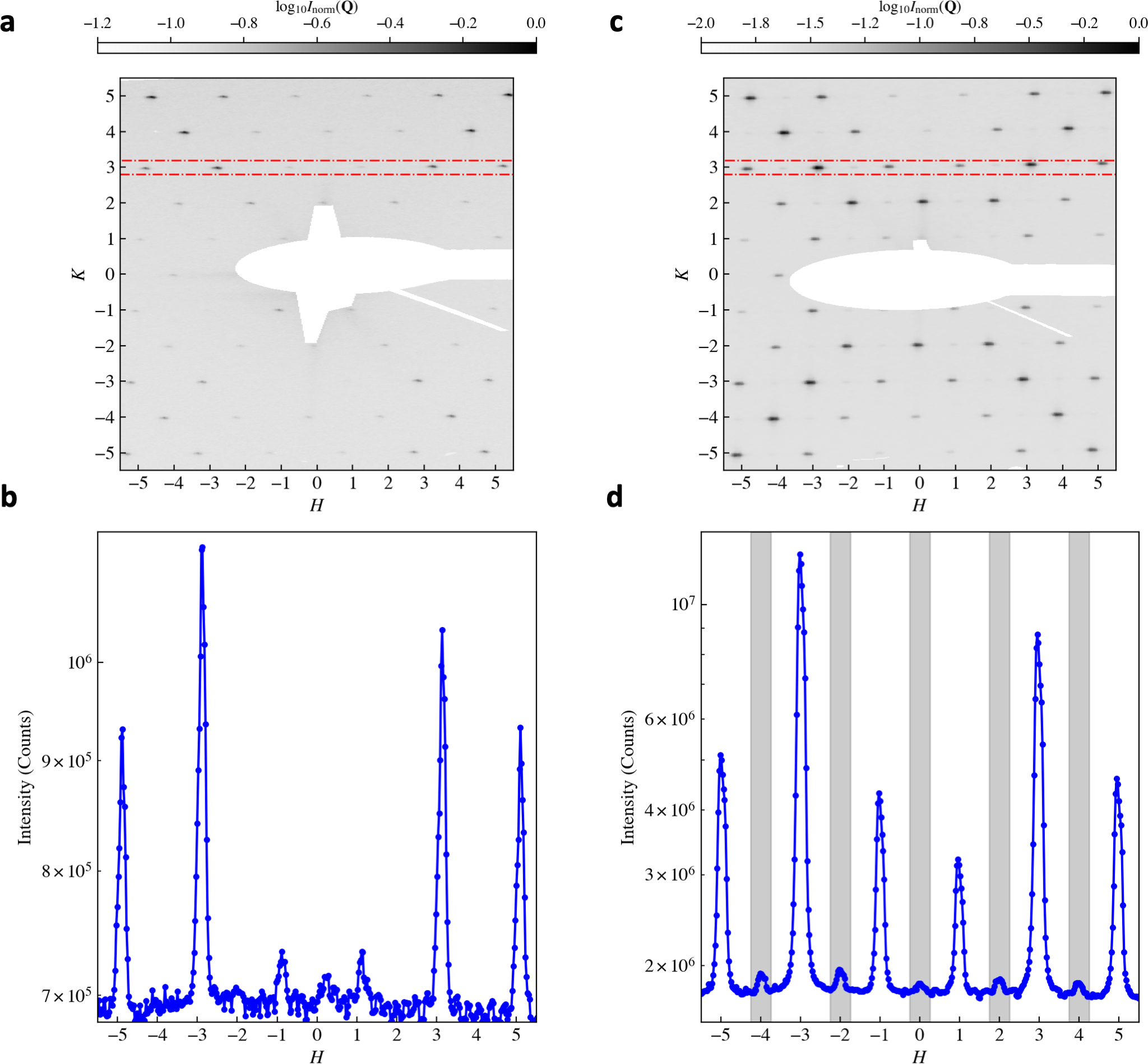}
\caption{\linespread{1.0} \footnotesize {\bf Off- and on-resonance scattering.} {\bf a} Diffraction pattern of the implanted ASI lattice, shown after corrections for the Ewald sphere curvature. Only charge scattering contributes at this energy, resulting in weak peak intensities and a narrow dynamic range due to the low electron density contrast between Fe-implanted regions and the Pd matrix. The characteristic “$\times$”-shaped envelope reflects the structural form factor of the mesospin basis. The white shaded area corresponds to the beam stop covering the specular reflection. {\bf b} The integrated intensity along $K$ as a function of $H$ for a region of interest centred on $K=3$ and marked with the red dash-dotted line in panel a. {\bf c} Resonant enhancement increases the structural Bragg peak intensity and reveals additional reflections associated with long-range antiferromagnetic order. The clearer “$\times$”-shaped modulation highlights the coherent mesospin basis form factor, while the emergence of mixed-parity peaks evidences the magnetic contribution to the scattering. {\bf d} The integrated intensity as a function of $H$ for a region of interest centred on $K=3$ and marked with the red dash-dotted line in panel c. Gray-shaded areas denote the positions of the reflections arising due to the antiferromagnetic order on the ASI lattice (Fig. \ref{PEEM-XMCD}a).} 
\label{off-on-resonance}
\end{figure}

A geometric interpretation of the diffraction condition is provided by the Ewald construction (see Fig.~\ref{Ewald}). In our experiment, diffraction patterns were recorded as two-dimensional CCD images at fixed incidence angle ($\theta=\ang{15}$), which sample a curved reciprocal-space slice determined by the Ewald sphere and detector plane. We did not perform coupled $\theta$--$2\theta$ rocking scans to integrate the full truncation-rod intensity. As a result, diffracted intensity is recorded at discrete pixel locations on the detector (Supplementary Fig.~\ref{Detector}). By combining the diffractometer geometry—including the sample-to-detector distance and angular coordinates—with the pixel positions, the detector coordinates were converted into the laboratory-frame momentum transfer vectors $\mathbf{Q}$. With additional knowledge of the sample periodicity, a further transformation was then applied to express the scattering data in reciprocal lattice units, enabling direct identification of the corresponding $(H,K)$ indices of the array’s reciprocal lattice (see Supplementary Information). After this conversion, any shifts of the recorded peak positions away from integer values of $H$ or $K$ can be attributed to uncertainties in the diffractometer angles, sample-to-detector distance, or sample orientation, all of which influence the final transformation to reciprocal lattice units  \cite{Schleputz_Q_conversion2011}.

A representative diffraction pattern from the implanted array is shown in Fig.~\ref{off-on-resonance}a. A prominent feature is the abundance of diffraction peaks, which occur at positions where both $H$ and $K$ are either simultaneously even or odd. These intensity peaks reflect the structural in-plane periodicity. Although at this energy the scattering contrast between the Pd matrix and the Fe-implanted regions is relatively low, resulting in only modest peak intensities, the diffraction peaks remain sharp, indicating long-range structural coherence across the array. The diffraction pattern intensities exhibit mirror symmetry with respect to the $H$ axis but not the $K$ axis. This asymmetry in $K$ originates from the geometric coupling between the out-of-plane momentum transfer $Q_z$ and the in-plane component $Q_y$, which arises from the curvature of the Ewald sphere projection onto the flat detector. The corresponding line scan, integrated region of interest centered on $K = 3$ (red dashed-dot rectangles in Fig.~\ref{off-on-resonance}a),  further confirms the modest peak intensities, their sharpness, and the absence of any half-order reflections. Immediately apparent in the line scan is the mirror symmetry with respect to $H$, as well as a pronounced and non-trivial dependence of the peak intensities with the diffraction order.

\begin{figure}[t]
\includegraphics[width=8cm]{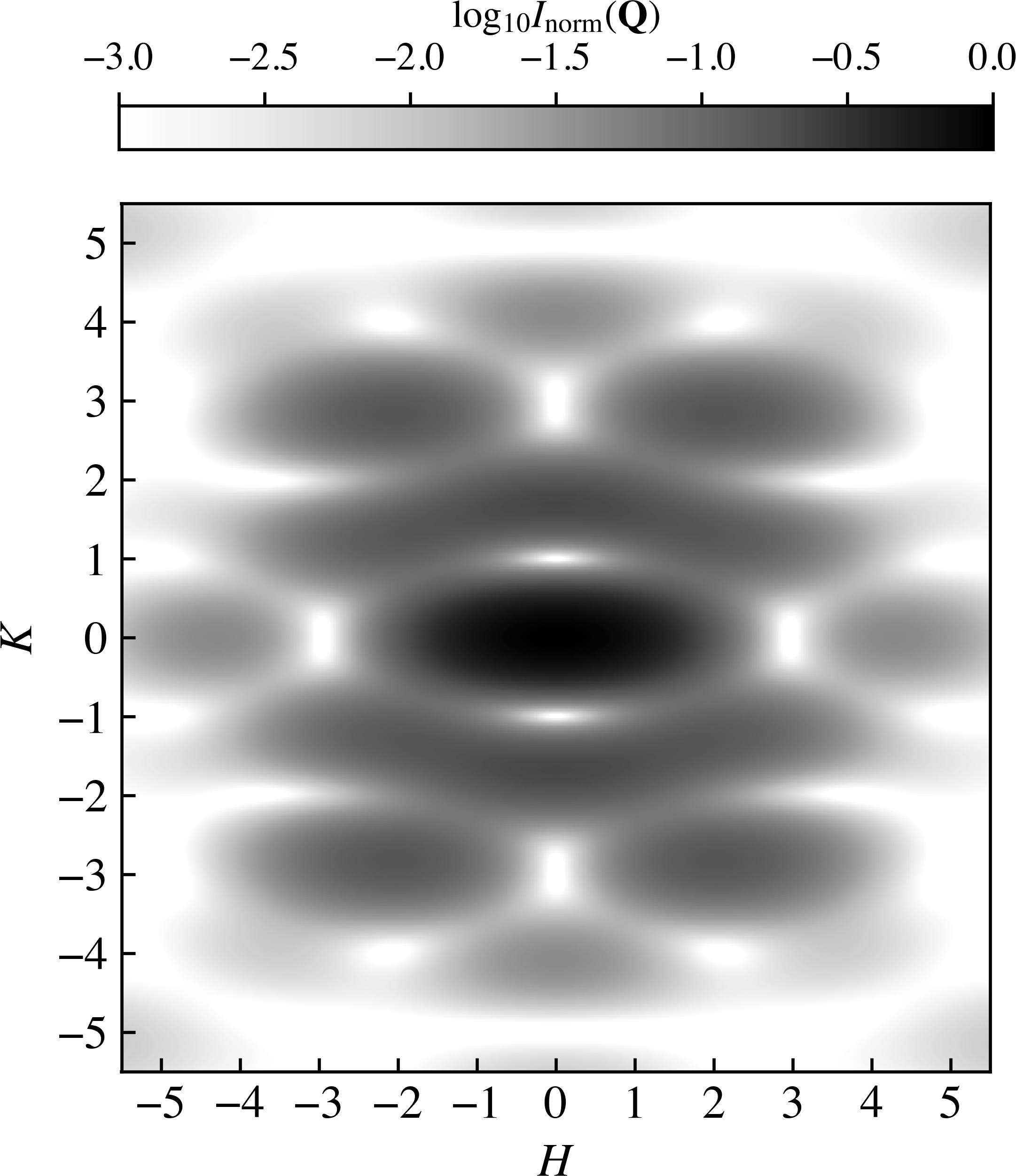}
\caption{\linespread{1.0} \footnotesize {\bf Mesospin form factor.} Calculated structural form factor $|F(\mathbf{Q})|$ for the four-mesospin basis forming a square-ASI vertex, using stadium-shaped mesospins as defined in the sample details ($L=\SI{470}{nm}$, $W=\SI{170}{nm}$). The reciprocal-space axes are expressed in ASI lattice units.}
\label{form-factor}
\end{figure}

A striking feature observed in Fig.~\ref{off-on-resonance}a is the characteristic “×”-shaped distribution of scattered intensity centred at $(0,0)$. This pattern can be understood within the framework of the convolution theorem, by which the diffracted amplitudes from a crystal are given by the Fourier transform of the real-space lattice multiplied by the Fourier transform of the basis. The measured intensity is then the amplitude squared. In the present case, the basis consists of two stadium-shaped mesospins within each unit cell, defined by their specific spatial arrangement and orientation \cite{alma991018439175107596}. To model this contribution, each mesospin was represented by a stadium geometry consistent with the sample details: a rectangle of length $L-W$ and width $W$, capped by semicircles of diameter $W$, with $L=\SI{470}{nm}$ and $W=\SI{170}{nm}$. The square of the calculated Fourier transform of this basis, expressed in reciprocal lattice units, is shown in Fig.~\ref{form-factor}. The intensity distribution exhibits enhanced scattering along the diagonal directions, consistent with the experimentally observed patterns. This demonstrates how the morphology and arrangement of the array elements modulate the scattering envelope and imprint a directional anisotropy of the intensity in reciprocal space. This effect is further amplified at resonance due to the increased scattering contrast and the concomitant enhanced visibility of the basis itself. Notably, such a clear manifestation of the basis form factor has not been reported in lithographically patterned square ASI lattices \cite{ChenX.M.2019SMSW,PerronJ.2013Erso,McCarterMargaretR.2023Arcp}, where fabrication-induced variations in element shape and size may obscure interference effects associated with the basis geometry \cite{DigernesEinar2020Diol}. In contrast, ion implantation, as reported here, produces more morphologically uniform mesospins, yielding a better and well-defined scattering envelope.

Tuning the X-ray energy to the Fe $L_3$ absorption edge significantly alters the measured diffraction pattern, as shown in Fig.~\ref{off-on-resonance}c. Together with the Fe $L_{2,3}$ absorption spectrum (Supplementary Fig.~\ref{SI_XAS_5deg}), this on/off-resonance contrast directly confirms Fe sensitivity: at resonance the total counts increase and magnetic peaks emerge, whereas off resonance they are absent. The resulting magnetic Bragg peaks, which arise from the long-range antiferromagnetic order of the islands, appear at reciprocal lattice positions where $H$ and $K$ are mixed odd and even (see Supplementary Information). A closer inspection of the line scans in Fig.~\ref{off-on-resonance} reveals the markedly different evolution of charge and magnetic peak intensities when moving on and off resonance. Off resonance (Fig.~\ref{off-on-resonance}b), only the structural reflections are present and their relative intensities follow the envelope imposed by the mesospin form factor, with no features at mixed-parity positions. Upon tuning to the Fe $L_3$ edge (Fig.~\ref{off-on-resonance}d), the structural peaks increase moderately due to the enhanced contrast between Fe-rich and Fe-poor regions, whereas the mixed-parity reflections appear exclusively at resonance and with intensities that scale with the magnetic contribution to the scattering factor. These magnetic peaks remain weaker than neighbouring structural peaks, consistent with the smaller magnetic scattering cross section and with the geometry of the experiment, which selects only the magnetization components parallel to $Q_y$. The systematic absence of these reflections off resonance, combined with their appearance at the symmetry-expected positions and their intensity scaling, unambiguously identifies them as magnetic Bragg peaks rather than artefacts of higher harmonics or multiple scattering. The relative charge–to–magnetic peak ratios are also in qualitative agreement with the kinematic simulations discussed below, which include both the mesospin form factor and the probe sensitivity axis, supporting their assignment to the Type-I antiferromagnetic ground state.

\begin{figure}[t]
    \centering
    \includegraphics[width=8cm]{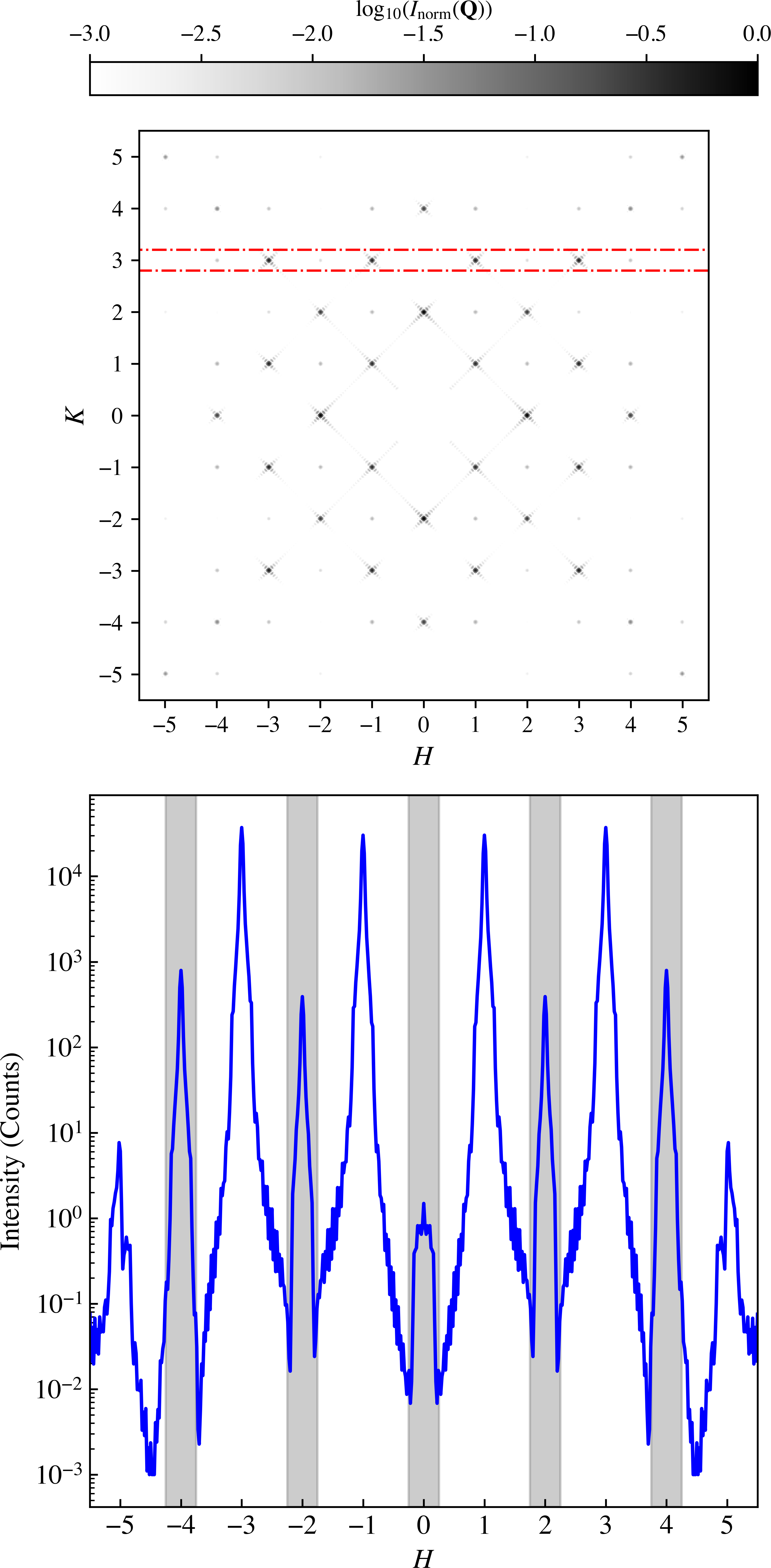}
    \caption{\linespread{1.0} \footnotesize {\bf Simulated line scans for the Type-I antiferromagnetic ground state.} Calculated intensity profiles, comparable to Fig.~\ref{off-on-resonance}b,d (along $H$ around $K\approx 3$), include structural and magnetic contributions together with the mesospin form factor and X-ray sensitivity axis, and capture the overall experimental intensity pattern.}
    \label{ssf_synth}
\end{figure}

Finally, for completeness, Fig.~\ref{ssf_synth} presents simulated line scans from a Type-I tiling of the islands, extracted in a geometry comparable to Fig.~\ref{off-on-resonance}b,d. The model combines the calculated structural and magnetic contributions, the mesospin form factor, and the sensitivity axis of the incident X-rays. The contribution from the form factor is accounted for as:

\begin{equation}
I(\mathbf{Q}) = |A(\mathbf{Q})|^2 = |S(\mathbf{Q})|^2 \, |F(\mathbf{Q})|^2 ,
\label{form_factor_I}
\end{equation}

\noindent Here $S(\mathbf{Q})=\sum_{j=1}^{N} s_j e^{i\mathbf{Q}\cdot\mathbf{r}_j}$ is the complex lattice-sum (structure-factor amplitude), and $|S(\mathbf{Q})|^2$ is the corresponding intensity structure factor, while $F(\mathbf{Q})$ is the mesospin form factor \cite{Als-Nielsen_Book}. It is worth noting that, in the most general case, one should account for two distinct form factors: a structural form factor, reflecting the Fe implantation profiles, and a magnetic form factor, describing the magnetization texture within each mesospin. The resulting line scans of Fig. \ref{ssf_synth} reproduce the overall intensity modulation observed experimentally. A detailed one-to-one comparison, however, would require additional modelling beyond this kinematic treatment, including the Ewald-sphere projection onto the planar detector, the intensity profile along the truncation rods sampled by the fixed-angle geometry, and instrumental broadening/resolution effects. 

Both experimental and simulated diffraction patterns reveal a pronounced $\mathbf{Q}$ modulation of the scattered intensity. This modulation reflects the mesospin form factor and the sensitivity of resonant scattering to specific magnetization components. This offers a direct link between local geometry and global diffraction symmetry. The island shape and sensitivity axis of the technique lead to systematic absences and a non-trivial intensity dependence on diffraction order. Careful selection of $\mathbf{Q}$ values of interest is, therefore, crucial for identifying the most appropriate experimental geometry. Such an approach allows for targeted data acquisition that captures length-scale information directly associated with specific $\mathbf{Q}$ values and access information that is often underutilized in ASI studies. Historically, analyses of scattered intensities in ASIs have tended to be qualitative or restricted to a limited set of reflections accessible in default measurement geometries, often overlooking features that encode key information about magnetic correlations at other relevant length scales \cite{Sendetskyi_2016, Morley_2017, ChenX.M.2019SMSW, Singh_Nat_Comm_2023}. Expanding access to a broader region of reciprocal space, combined with kinematic scattering simulations (structure-factor modelling) and a clear link to the reciprocal lattice of the sample itself will enable future experiments to fully exploit the temporal and spatial coherence of resonant soft X-ray scattering. This advancement, in turn, promises unprecedented insight into collective ordering and dynamical processes in ASIs and magnetic metamaterials more generally. It also allows going beyond static snapshots, toward a deeper and more quantitative understanding, enabled by the more consistent and uniform mesospins produced by ion implantation.

\section{\label{sec:conclusions} Conclusions and Outlook}

Our combined use of resonant X-ray and neutron scattering, together with magnetic microscopy, reveals that ion implantation enables the formation of magnetically ordered metamaterials with structural and magnetic coherence extending over large areas. The implanted Fe ions generate a confined ferromagnetic profile within the Pd host by inducing polarization in the surrounding matrix. This localized magnetic region forms the building blocks of a class of {\it embedded mesospins}, whose coupling and geometry drive the spontaneous emergence of collective order.

Magnetic imaging confirms that each mesospin acts as a single-domain element and that the implanted islands organize into extended antiferromagnetic domains corresponding to the Type-I ground state of the square geometry. This long-range order emerges directly during the implantation process, without the need for post-fabrication annealing or magnetic field cycling. Our observations suggest that the local energy input, Fe$^+$ ion dose accumulation, and diffusion dynamics during implantation effectively promote self-thermalization, allowing the system to explore low-energy configurations as magnetic moments build up within the mesospins.

In reciprocal space, sharp structural and magnetic Bragg peaks testify to the high morphological uniformity of the implanted array. The scattering envelope obtained demonstrates that implantation overcomes the disorder and variability inherent to lithographic patterning, enabling access to quantitatively interpretable structure–factor regimes that might have been previously inaccessible in artificial spin systems. Comprehensive modeling provides valuable insight that will be instrumental in future sample design and experimental strategy, enabling optimized use of available scattering geometries and detector configurations. 

These results establish ion implantation as a scalable, additive approach to realize self-organized magnetic order in structurally coherent metamaterials. The coupling between magnetic and morphological coherence provides a pathway toward functional platforms for spin-resolved and photon-mediated phenomena, including spin--photon coupling, orbital-angular-momentum scattering, and magneto-optical interference in structured media \cite{OAM_ASI_PRL, McCarterMargaretR.2023Arcp}. More broadly, the spontaneous emergence of the ground state in an as-fabricated system highlights a potentially useful paradigm in artificial magnetism: magnetic metamaterials that evolve toward their ordered configurations during synthesis. Furthermore, ion implantation offers a unique avenue for directly engineering both structural and functional properties. Through careful selection of ion species, implantation energy and fluence, it becomes possible to tailor local magnetic anisotropy, ordering temperature, and moment magnitude \cite{StromPetter2022Soft, VantarakiChristina2024Mmbi, Vantaraki_inhomogeneities}---parameters that are typically less independently tunable in lithographically defined systems. Intriguingly, one could envisage extending this concept toward multi-ion or spatially graded 3D architectures, where controlled implantation profiles generate programmable magnetic functionalities or coupled responses across multiple length scales and dimensions. This capability establishes ion implantation not merely as a fabrication route but as a materials design tool for spatially resolved control of emergent magnetic and electronic behavior. Such systems have the potential to bridge the gap between materials design and dynamic self-organization, offering different opportunities for reconfigurable logic, magnonic information processing \cite{MR_RC_ASIs_2023, Gartside_RC, Gliders_Folven_2025}, and optical or scattering-based read-out schemes \cite{Opt_RC_IEEE_2019, Opt_RC_LSA_2025} in unconventional computing architectures \cite{Folven_computation, Penty_IJUC_2023}.

\section{Methods} \label{sec:methods}

\subsection{\label{subsec:samples} Sample fabrication} 

Initially, a continuous Pd film was deposited by DC magnetron sputtering onto a MgO substrate, with a \SI{5}{nm} V adhesion layer and a \SI{6}{nm} Cr capping layer that is in place to avoid ablation of the underlying Pd during ion implanting. Subsequently, a square ASI lattice was fabricated by implanting \SI{30}{keV} $^{56}$Fe$^{+}$ ions with a nominal fluence of $4 \times 10^{16}$~ions/cm$^{2}$ into a \SI{60}{nm} Pd film through a patterned Cr implantation mask that was subsequently removed using a wet etch. The Cr mask was defined lithographically prior to implantation; however, the magnetic elements themselves are not topographically patterned. This additive fabrication process results in Fe$_{x}$Pd$_{100-x}$ (where $x$ is in atomic percent) ferromagnetic structures embedded within the otherwise non-magnetic Pd matrix. The implanted Fe exhibits both lateral and vertical concentration profiles \cite{Vantaraki_inhomogeneities}, the extent of which is controlled through the patterned Cr mask and the implantation energy. After mask removal, the sample surface remains nearly planar; this suppresses the edge-roughness and thickness-variation disorder typical of free-standing lithographic islands while retaining lithographic control of lateral placement. A detailed description of the fabrication process and material properties is provided by Vantaraki {\it et al.} \cite{VantarakiChristina2024Mmbi, Vantaraki_inhomogeneities}. The particular ASI array fabricated for this study consists of stadium-shaped elements with dimensions of \SI{470}{nm} in length and \SI{170}{nm} in width, with an edge-to-edge gap of $g=$\SI{170}{nm}. A representative SEM image of the array is shown in Fig.~\ref{Figure1}.

In addition to the patterned film, two continuous Pd films implanted homogeneously with Fe ions were fabricated. The Pd layers had a nominal thickness of either \SI{40}{nm} or \SI{60}{nm} and were deposited by DC magnetron sputtering onto MgO substrates, each with a \SI{5}{nm} V adhesion layer and a \SI{6}{nm} Cr capping layer. Growth and processing conditions were identical to the films used to create the arrays. The \SI{40}{nm} film was deposited on a $10\times$ \SI{10}{\mm\squared} substrate, whilst the \SI{60}{nm} film was deposited on a $20 \times$ \SI{20}{\mm\squared} substrate to facilitate polarized neutron reflectometry measurements. Ion implantation of the continuous films was carried out under the same conditions as for the patterned array, using \SI{30}{keV} $^{56}$Fe$^{+}$ ions at a nominal fluence of $4 \times 10^{16}$~ions/cm$^{2}$. Owing to their extended geometry, the implanted Fe concentration in these films exhibits only a depth-dependent profile. \cite{StromPetter2022Soft}.

\subsection{\label{subsec:magn_prof} Magnetization depth profiling} 

The element-specific Fe magnetization of the implanted samples was determined using resonant synchrotron reflectivity measurements at the SEXTANTS beamline at the SOLEIL synchrotron \cite{SacchiM2013TSba}. The experiments were performed in a reflection geometry on the \SI{40}{nm} continuous Fe$^{+}$-implanted Pd film at room temperature, with the magnetization saturated along the X-ray sensitivity axis with an applied magnetic field of 50 mT. The scattered intensity was recorded for left ($I^{L}$) and right ($I^{R}$) circularly polarized light, with the photon energy tuned to the Fe $L_{3}$ edge (\SI{707}{eV}). The resonance energy selection is supported by the Fe $L_{2,3}$ X-ray absorption spectrum shown in Supplementary Information. The sum of the two reflectivity curves, $I^{L} + I^{R}$, and the corresponding asymmetry ratio, $AR = (I^{L} - I^{R})/(I^{L} + I^{R})$, were simultaneously fitted using the \textsc{GenX} software package to extract both the chemical and magnetic depth profiles through their respective scattering length densities \cite{BjörckMatts2007GaeX, Glavic_GenX3}. Complementary magnetization profiling studies were performed using Polarized Neutron Reflectometry (PNR) with one-dimensional spin analysis at the D17 beamline in ILL, Grenoble. Details of the latter are presented in depth in the Supplementary Information.

\subsection{Resonant X-ray scattering and microscopy} \label{subsec:scatt_microscopy}

The magnetic configuration of the implanted square ASI lattice was determined in real space using photoemission electron microscopy combined with X-ray magnetic circular dichroism (PEEM-XMCD), with measurements carried out at the CIRCE (BL24) beamline of the ALBA synchrotron \cite{AballeLucia2015TAsL}. The PEEM images were collected at room temperature and in the absence of any external magnetic field. The photon energy was tuned to the Fe $L_{3}$ edge to provide element-specific magnetic contrast. The stadium-shaped elements were oriented at \ang{45} with respect to the incident X-ray beam, allowing unambiguous determination of the magnetization direction for all elements in the lattice within the field of view. The magnetic state was imaged in the as-implanted, virgin configuration.

The lateral magnetic ordering of the implanted ASI lattice was further examined in reciprocal space using soft X-ray scattering experiments conducted at the SEXTANTS beamline of the SOLEIL synchrotron \cite{SacchiM2013TSba}. The measurements were carried out in reflection geometry, with diffraction patterns from the periodic lattice captured using a Charge Coupled Device (CCD) detector. To shield the detector from the intense specularly reflected beam, a beam stop was deployed. A circularly polarized X-ray beam, tuned to the Fe L$_{3}$ edge, provided element-specific sensitivity to the magnetization of the mesospins. The angle of incidence, $\theta$, was set to \ang{15} to enhance sensitivity to the in-plane magnetization. Consistent with the microscopy measurements, the ASI lattice was aligned so that the X-ray beam propagated along the [1,1] direction of the square array.

\subsection{Kinematic scattering simulations}

We performed kinematic scattering simulations of the square ASI by constructing a finite Type-I antiferromagnetic lattice in the experimental geometry, with the incident beam aligned along the [1,1] direction of the array. The charge channel was calculated from the coherent sum of all islands, weighted by a structural basis form factor $F(\mathbf{Q})$ for the two orthogonal stadium-shaped mesospins in each unit cell, using the sample dimensions $L=\SI{470}{nm}$ and $W=\SI{170}{nm}$. The magnetic channel was computed separately from the ordered mesospin moments using the probe sensitivity axis appropriate to the resonant soft X-ray geometry, and the total simulated intensity was obtained by combining the structural and magnetic contributions. Reciprocal-space maps and line cuts were then extracted in the same $(H,K)$ indexing used for the experimental analysis, with optional masking of the central beam-stop region for direct comparison with the measured diffraction patterns.


\subsection{Acknowledgments}
\footnotesize
The authors would like to thank Johan Oscarsson and Mauricio Sortica at the Uppsala Tandem Laboratory for help with ion implantations. CV and VK would like to thank Prof. Bj\"orgvin Hj\"orvarsson for fruitful discussions. The authors would like to thank Prof. Bengt Lindgren for fruitful discussions on fitting the resonant X-ray scattering and PNR data.

\subsection{Funding declaration}
\footnotesize
The authors are thankful for an infrastructure grant by VR-RFI (grant number 2019-00191) supporting the accelerator operation. The authors also acknowledge support from the Swedish Research Council (projects no. 2019-03581 and 2023-06359). CV gratefully acknowledges financial support from the Colonias-Jansson Foundation, Thelin-Gertrud Foundation, Liljewalch and Sederholm Foundation. OB acknowledges support from EPSRC through the XMaS National Research Facility at the ESRF. We acknowledge Myfab Uppsala for providing facilities and experimental support. Myfab is funded by the Swedish Research Council (2020-00207) as a national research infrastructure. The PEEM-XMCD experiments were performed at CIRCE (bl24) beamline at ALBA Synchrotron with the collaboration of ALBA staff. MF and MANO acknowledge support from MICIN through grant number PID2021-122980OB-C54. Funded by the European Union as part of the Horizon Europe call HORIZON-INFRA-2021-SERV-01 under grant agreement number 101058414 and co-funded by UK Research and Innovation (UKRI) under the UK government’s Horizon Europe funding guarantee (grant number 10039728) and by the Swiss State Secretariat for Education, Research and Innovation (SERI) under contract number 22.00187.

Views and opinions expressed are, however, those of the authors only and do not necessarily reflect those of the European Union or the UK Science and Technology Facilities Council or the Swiss State Secretariat for Education, Research and Innovation (SERI). Neither the European Union nor the granting authorities can be held responsible for them.

\subsection{Author contributions}
CV, TPHA and VK conceived the project. CV, DP and VK developed the sample fabrication processing, combining electron-beam lithography with ion implantation. CV, OB, MPG, BO, AS, NJ, TPAH and VK performed all the resonant X-ray scattering measurements. CV, TS, MW and VK performed the neutron scattering experiments. CV, MPG, N K-M, MANO and MF performed the magnetic microscopy studies. Data analysis was performed primarily by CV, OB, MPG, TPAH and VK with input from all authors. TPHA and VK wrote the manuscript. All authors discussed the results and commented on the manuscript. VK provided funding and supervised the project.

\subsection{Competing financial interests}
The authors declare no competing financial interests.

\subsection{Data availability}

The data that support the findings of this study are available from the corresponding authors upon reasonable request. Neutron scattering data are available at: KAPAKLIS VASSILIOS, CUBITT Robert, SAERBECK Thomas, VANTARAKI Christina, \& WOLFF Maximilian. (2024). Magnetic metamaterials produced by ion-implantation (ReMade - TNA) [Dataset]. Institut Laue-Langevin (ILL). DOI:\href{https://doi.org/10.5291/ILL-DATA.DIR-329}{10.5291/ILL-DATA.DIR-329}.

\normalsize

\newpage
\onecolumngrid
\begin{center}
\textbf{\large Supplementary Information:\\ Long-Range Structural and Magnetic Coherence \\ in Embedded Mesospin Metamaterials}

\end{center}

\begin{outlinesec}
    \section{Supplementary Information}
\end{outlinesec}

\setcounter{equation}{0}
\setcounter{figure}{0}
\setcounter{table}{0}
\setcounter{page}{1}
\makeatletter
\renewcommand{\theequation}{S\arabic{equation}}
\renewcommand{\figurename}{{\bf Supplementary Fig.}}
\renewcommand{\thefigure}{{\bf S\arabic{figure}}}
\renewcommand{\bibnumfmt}[1]{[S#1]}
\renewcommand{\citenumfont}[1]{S#1}
\renewcommand{\thepage}{S-\arabic{page}}
\renewcommand{\tablename}{{\bf Supplementary Table}}
\renewcommand{\thetable}{{\bf S-\arabic{table}}}

\subsection{\label{Detector_raw} X-ray scattering detector images}

\begin{figure}[ht!]
 \centering
    \includegraphics[width=8cm]{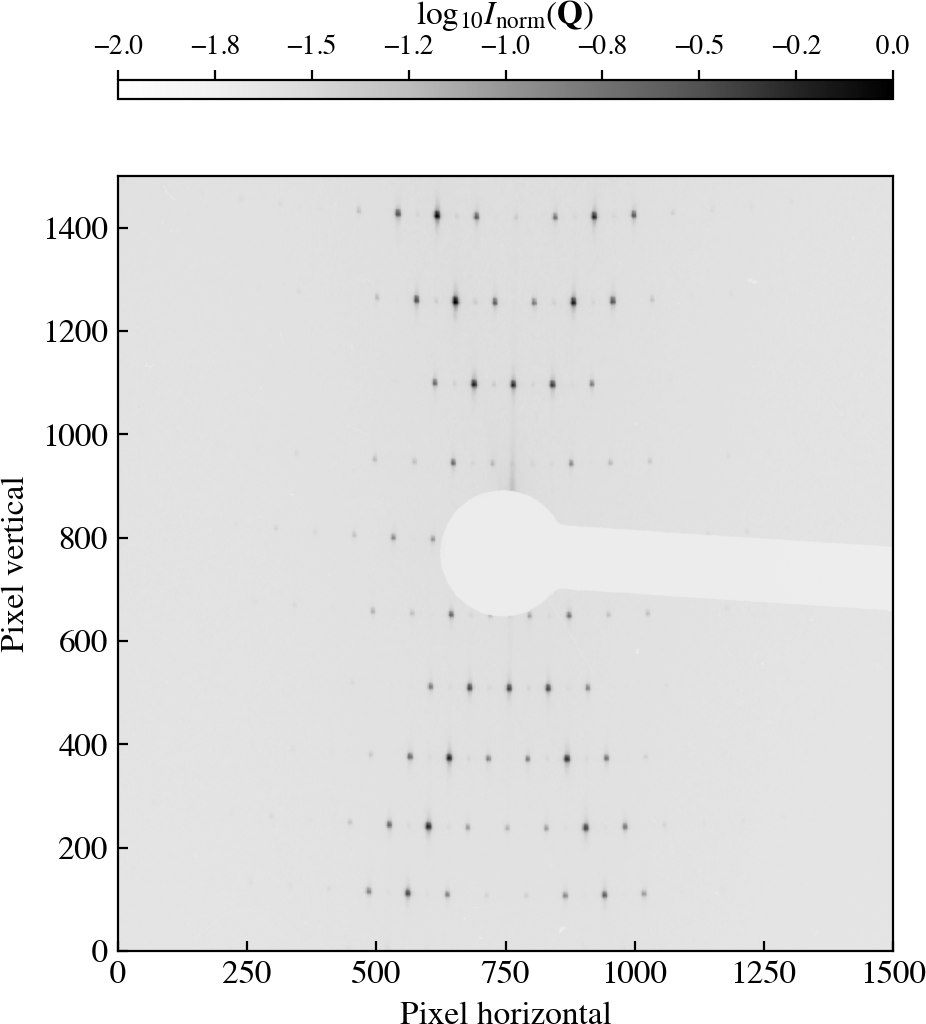}
    \caption{\linespread{1.0} \footnotesize {\bf Raw CCD detector image acquired on-resonance.} Unprocessed pixel data from a resonant soft X-ray scattering snapshot. The apparent curvature of the diffraction features arises because the 2D detector captures a curved slice through reciprocal space corresponding to the portion of the Ewald sphere intersecting the Bragg rods.}
    \label{Detector}
\end{figure}

The raw CCD detector images (Supplementary Fig.~\ref{Detector}) already reveal the key features expected for scattering from a square-lattice array under the given experimental geometry. Because the mesospin pitch is on the order of several hundred nanometres, the corresponding diffraction peaks appear close to the specular reflection, which in our case is masked by a beam stop. As the detector samples a curved section of reciprocal space—specifically, the set of reciprocal-space points lying on the portion of the Ewald sphere intercepted or projected onto the detector \cite{Schleputz_Q_conversion2011_2} (Fig.~\ref{Ewald})—the raw images can appear distorted, with peaks following curved trajectories rather than the straight lines expected from a square lattice. 

A precise mapping of the pixel coordinates of the image to the corresponding ($H,K$) values in reciprocal space is essential for quantitative analysis. This mapping can be performed once the exact experimental geometry and beam parameters are known. Importantly, this mapping also compensates for the distortions introduced by the experimental geometry 

\subsection{\label{PNR} Polarized neutron reflectometry}

Polarized Neutron Reflectometry (PNR) with one-dimensional spin analysis was performed at the D17 beamline at ILL, Grenoble \cite{SaerbeckThomas2018Ruot, ILL_DOI} on a $20 \times$ \SI{20}{\mm\squared} film of 60 nm Pd implanted with 30 keV Fe$^{+}$, under a saturating magnetic field of 100 mT. Measuring the reflectivity of neutron spins parallel ($R^{+}$) and anti-parallel ($R^{-}$) to the magnetization of the sample allows to extract the magnetic depth profile. The $R^{+}$ and $R^{-}$ reflectivity curves, along with the spin asymmetry $SA=\frac{R^{+} - R^{-}}{R^{+} + R^{-}}$, were fitted simultaneously using the \textsc{GenX} program \cite{BjörckMatts2007GaeX_2, Glavic_GenX3_2}.

\begin{figure}[t]
\includegraphics[width=8cm]{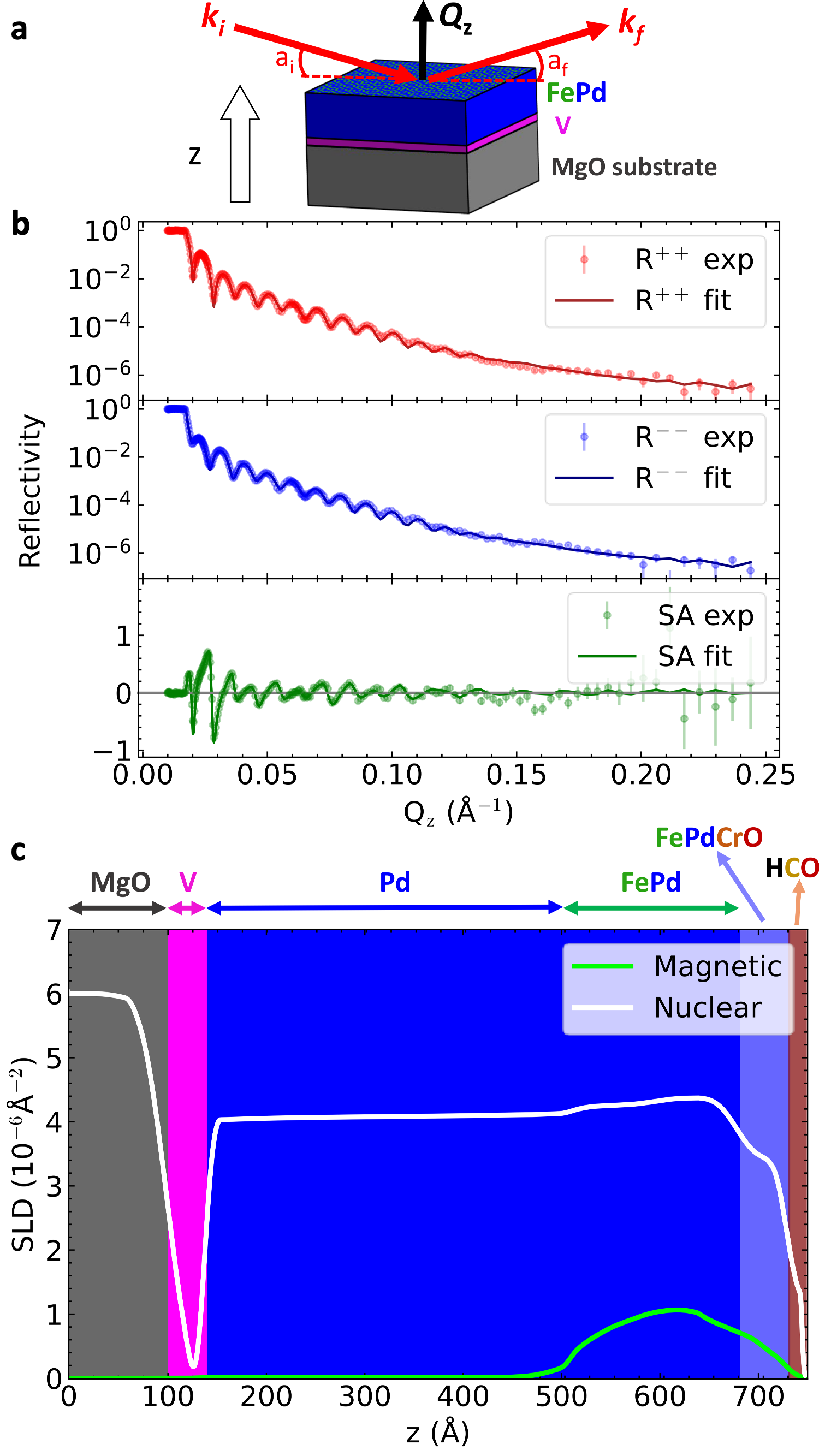}
\caption{\linespread{1.0} \footnotesize {\bf Polarized neutron reflectivity.} {\bf a} Schematic of the 30 keV Fe$^{+}$-implanted film. Polarized neutron reflectometry was performed in specular geometry ($a_i = a_f$), as a function of out-of-plane momentum transfer $Q_z$. {\bf b} Neutron reflectivity curves for spin-up ($R^{+}$) and spin-down ($R^{-}$) channels, and the spin asymmetry (SA). Symbols represent experimental data; solid lines correspond to the best-fit model. {\bf c} Derived nuclear and magnetic scattering length density (SLD) profiles. Background shading indicates the MgO substrate, V adhesion layer, Pd film and impurity regions.
}
\label{PNR-plot}
\end{figure}

Unlike the X-ray data, the magnetic depth profile obtained from PNR measures the distribution of magnetization in depth and includes contributions from both Fe and Pd. The PNR curves are shown in Supplementary Fig.~\ref{PNR-plot}~(b), separately recorded for the two spin channels, $R^{+}$ and $R^{-}$. For implanted films, where ion mixing produces interdiffused chemical and magnetic gradients, the fit output is most robustly represented by the SLD depth profiles; a decomposition into unique layer thickness and interfacial roughness values is not physically unique. The Spin Asymmetry (SA) is directly proportional to the magnitude of the magnetic moment in the plane of the sample and along the quantisation axis. When fitting, we account for some Cr being implanted into the film as a result of knock-on collisions from the capping layer, based on previous results \cite{Vantaraki_inhomogeneities_2}.

The neutron reflectivity with spins parallel ($R^{+}$) and antiparallel ($R^{-}$) is fitted to a structural and magnetic model of neutron scattering length densities as a function of depth using the GenX program. The magnetic SLD is proportional to the in-plane magnetization as a function of depth ($\rho_{\mathrm{m}} = 2.91044\times 10^{-12}\,$Å$\mathrm{^{-2}\,m\,A^{-1}}\times M_{\perp}[\mathrm{Am^{-1}}]$). The spin asymmetry SA, highlighting only the magnetic contribution to the reflectivity, has been fitted simultaneously. Whilst the magnetic profile appears broader, the magnetic moment remains confined to within approximately \SI{15}{nm} of the surface. Since PNR measures the total magnetic moment, any induced magnetization in the Pd layer extends the magnetic thickness beyond the Fe-implanted region itself. 


\subsection{\label{Appendix_B} Lattices and Indexation}

\begin{figure}[t]
\includegraphics[width=6.5cm]{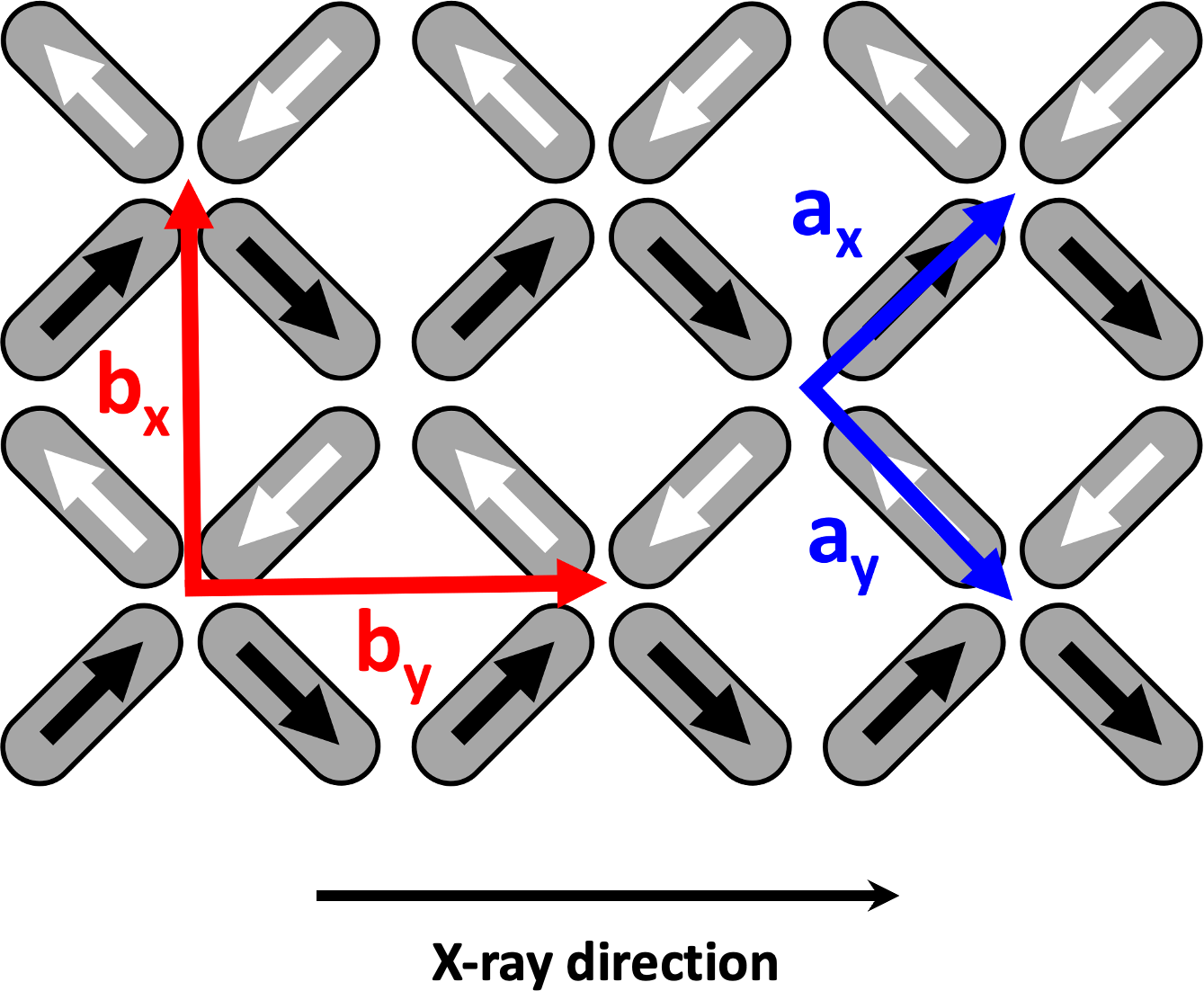}
\caption{\linespread{1.0} \footnotesize {\bf Real-space lattice vectors.} Real-space representation of the square ASI lattice and its magnetic ground state. The mesospin array forms a square Bravais lattice, while the Type-I antiferromagnetic configuration, combined with the X-ray sensitivity axis along the lattice diagonal, results in a magnetic unit cell enlarged by a factor of $\sqrt{2}$ and rotated by \ang{45}, i.e., a ($\sqrt{2},\times,\sqrt{2}$)R\ang{45} reconstruction. This defines the indexing scheme used to assign structural and magnetic Bragg peaks in reciprocal space.}
\label{real-space}
\end{figure}

A key step in interpreting the experimentally obtained scattering data is to index the diffracted intensities to a suitable lattice that describes the sample itself, rather than relying solely on the reciprocal space defined by the diffractometer axes. The islands themselves are arranged on a simple 2D square Bravais lattice denoted by the vectors $\vec{a}_{1,2}$ (Supplementary Fig.~\ref{real-space}). The reciprocal lattice vectors are given by:

\[ \vec{a}_{1}^{*}=2\pi\frac{\vec{a}_{2}\times\hat{n}}{\|A\|} \quad ; \quad \vec{a}_{2}^{*}=2\pi\frac{\hat{n}\times\vec{a}_{1}}{\|A\|}, \]

\noindent where $\hat{n}$ is the unit vector normal to the 2D plane and $\|A\|$ is the magnitude of the scalar triple product $\vec{a}_{1} \cdot \left(\vec{a}_{2}\times\hat{n}\right)$. A  translation in reciprocal space is given by $\vec{G}_{h,k} = h \vec{a}_{1}^{*} + k \vec{a}_{2}^{*}$.

When all vertices in the square ASI are of the Type-I configuration, the primitive magnetic lattice is $\sqrt{2}$ times larger than the Bravais lattice of the islands and is rotated \ang{45} with respect to it, Supplementary Fig.~\ref{real-space}. In Wood notation \cite{WOOD1964}, the magnetic lattice is $\left(\sqrt{2}\times\sqrt{2}\right)R\ang{45}$ or in matrix form\cite{PARK1968188}, $\begin{bsmallmatrix} 
1 & \bar{1}\\ 1 & 1 \end{bsmallmatrix}$.

\begin{figure}[t]
\includegraphics[width=0.7\linewidth]{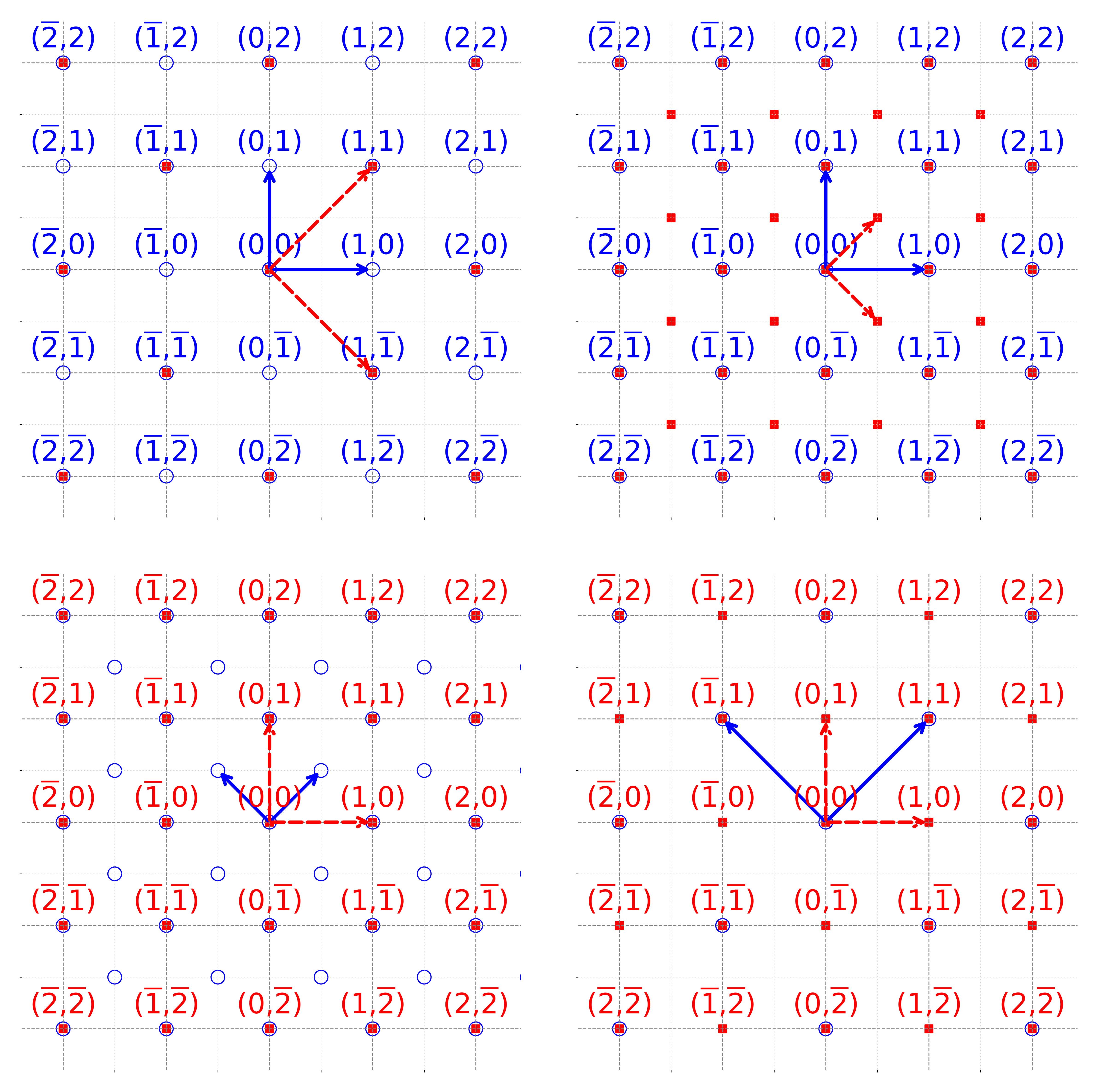}
\caption{\linespread{1.0} \footnotesize {\bf Reciprocal lattice indexing.} Representations of the array lattice in real (left) and reciprocal space (right). The array can be indexed to the simple square Bravais lattice associated with the islands (top) or the $\left(\sqrt{2}\times\sqrt{2}\right)R\ang{45}$ unit cell of the magnetic Type-I configuration (bottom).
}
\label{Lattices}
\end{figure}

To better visualise reciprocal space, and noting the \ang{45} rotation of the magnetic to island lattice with respect to the islands, we can re-index the lattice using a new set of vectors aligned along the [1,1] direction with basis vectors $\vec{b}_{1,2}$. Here  $\vec{b}_{1}$ is aligned with the [1,1] direction and $\vec{b}_{2}$ perpendicular to it, along the [1,$\bar{1}$] direction (Supplementary Figs.~\ref{real-space} and \ref{Lattices}). The relationship between the two indexations is simply $\vec{b}_{1} = \vec{a}_{1} + \vec{a}_{2}$ and $\vec{b}_{2} = \vec{a}_{1} - \vec{a}_{2}$. Normalising these vectors, i.e. $\hat{b}_{1,2} = \frac{1}{\sqrt{2}}\vec{b}_{1,2}$ simplifies the indexing to this new lattice description with the transformation, written in matrix form, as follows:

\begin{equation}
\begin{pmatrix}
\vec{b}_{1} \\
\vec{b}_{2}
\end{pmatrix}
= M
\begin{pmatrix}
\vec{a}_{1} \\
\vec{a}_{2}
\end{pmatrix}
= \frac{1}{\sqrt{2}}
\left(
\begin{array}{rr}
1 & \bar{1} \\
1 & 1
\end{array}
\right)
\begin{pmatrix}
\vec{a}_{1} \\
\vec{a}_{2}
\end{pmatrix},
\end{equation}

\noindent which in reciprocal space becomes:

\begin{equation}
\begin{aligned}
\begin{pmatrix}
\vec{b}_1^{*} \\
\vec{b}_2^{*}
\end{pmatrix}
&= M^{*}
\begin{pmatrix}
\vec{a}_1^{*} \\
\vec{a}_2^{*}
\end{pmatrix}
= \left(M^{-1}\right)^{\mathrm{T}}
\begin{pmatrix}
\vec{a}_1^{*} \\
\vec{a}_2^{*}
\end{pmatrix} \\
&= \frac{1}{\sqrt{2}}
\left(
\begin{array}{rr}
1 & 1 \\
\bar{1} & 1
\end{array}
\right)
\begin{pmatrix}
\vec{a}_1^{*} \\
\vec{a}_2^{*}
\end{pmatrix}.
\end{aligned}
\end{equation}

\noindent In this indexing scheme, as shown in Supplementary Fig.~\ref{Lattices}, structural diffraction from the islands is observed when both $H$ and $K$ are either even or odd. In contrast, magnetic Bragg peaks only appear at positions where $H$ and $K$ are mixed, i.e. one is even and the other is odd.

Under resonant X-ray scattering conditions, however, the magnetic scattering cross section becomes sensitive to the projection of the magnetic moment along the direction of the incident beam. This directional sensitivity modifies the observed symmetry of the magnetic order when the beam is incident along the [1,1] direction, effectively reducing the symmetry to a $2 \times 1$ construction (in terms of $\vec{b}$). Consequently, certain diffraction peaks become systematically absent in this geometry, depending on the parity of $H$ and $K$.

\subsection{\label{Appendix_C} Spin structure factor maps}

The long-range magnetic order observed in the PEEM images (Fig.~\ref{PEEM-XMCD}) provides an independent confirmation of the indexing procedure. The square modulus of the Fourier transform of the PEEM data, the spin structure factor (SSF), directly reveals the magnetic lattice in reciprocal space. The Fourier transform of the complete set of microscopy images is shown in Supplementary Fig.~\ref{Appendix_full_SSF} and can be indexed to the $\left(\sqrt{2}\times\sqrt{2}\right)R\ang{45}$ symmetry described above. The presence of sharp diffraction peaks further confirms the long-range magnetic order seen within the field of view of the PEEM instrumentation. Immediately apparent is the lack of any “×” feature (see main text) and no systematic absences. The magnetic configuration of the islands is fully described by the Type-I vertex configuration and its associated symmetry. The binary contrast in the intensities measured in the PEEM data explains the lack of any basis contribution.

\begin{figure}[t]
 \centering
    \includegraphics[width=7cm]{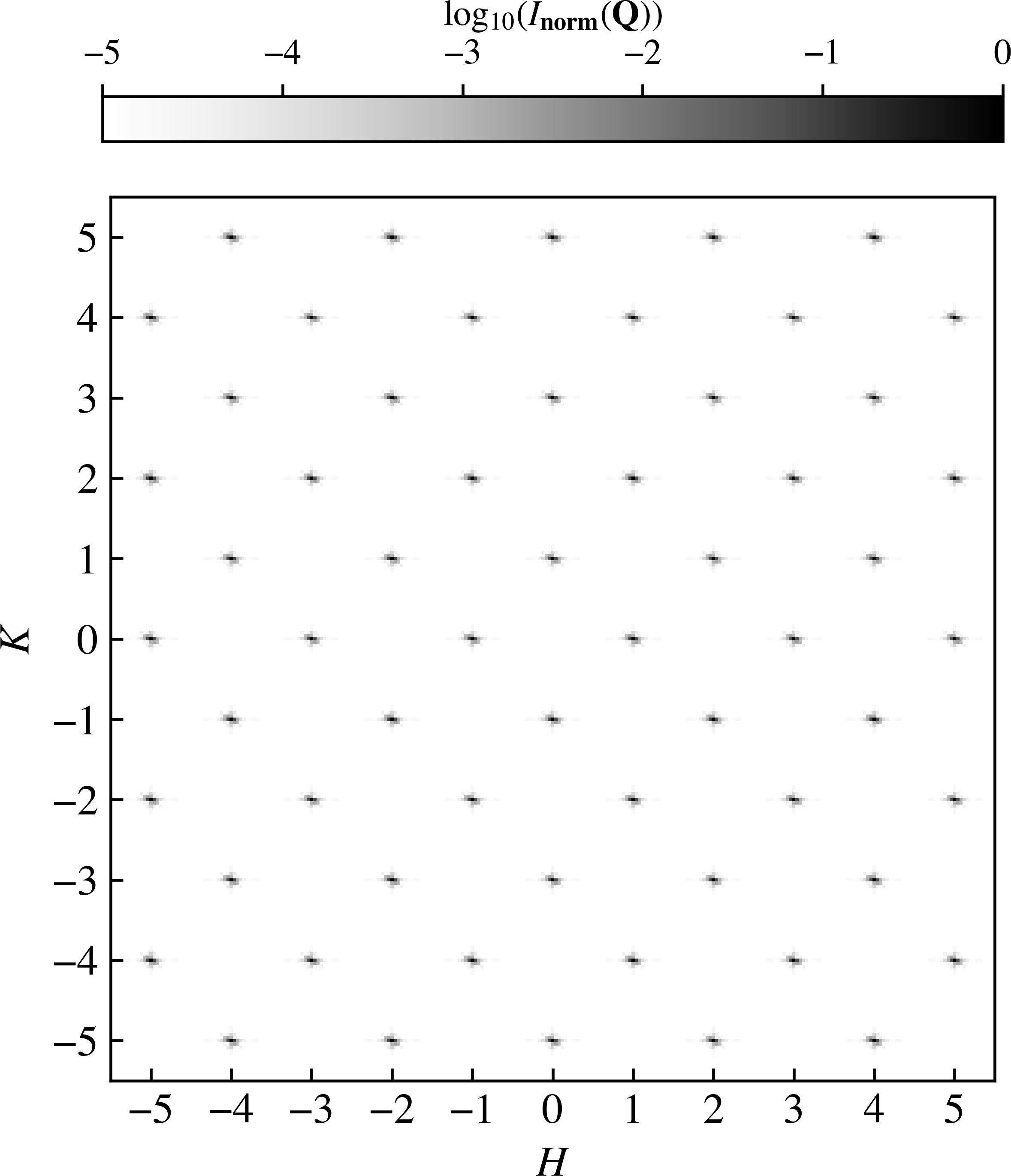}
    \caption{\linespread{1.0} \footnotesize {\bf The average SSF map calculated from the PEEM-XMCD images.} Sharp Bragg peaks are seen due to the extended antiferromagnetic order.}
    \label{Appendix_full_SSF}
\end{figure}

Given the experimental evidence for Type-I vertex configurations, we simulate the SSF maps by considering an infinite lattice of mesospin islands with spin $S$. The scattered intensity is obtained from the square of the Fourier transform of the spin–spin correlations,

\begin{equation}
I(\mathbf{Q}) = \frac{1}{N^{2}} \left| \sum_{i,j=1}^{N} \mathbf{S}_{i}\mathbf{S}_{j}  e^{i\mathbf{Q}\cdot(\mathbf{r}{i}-\mathbf{r}{j})} \right|^{2}.
\label{SSF_par}
\end{equation}

\begin{figure}[ht!]
 \centering
    \includegraphics[width=7cm]{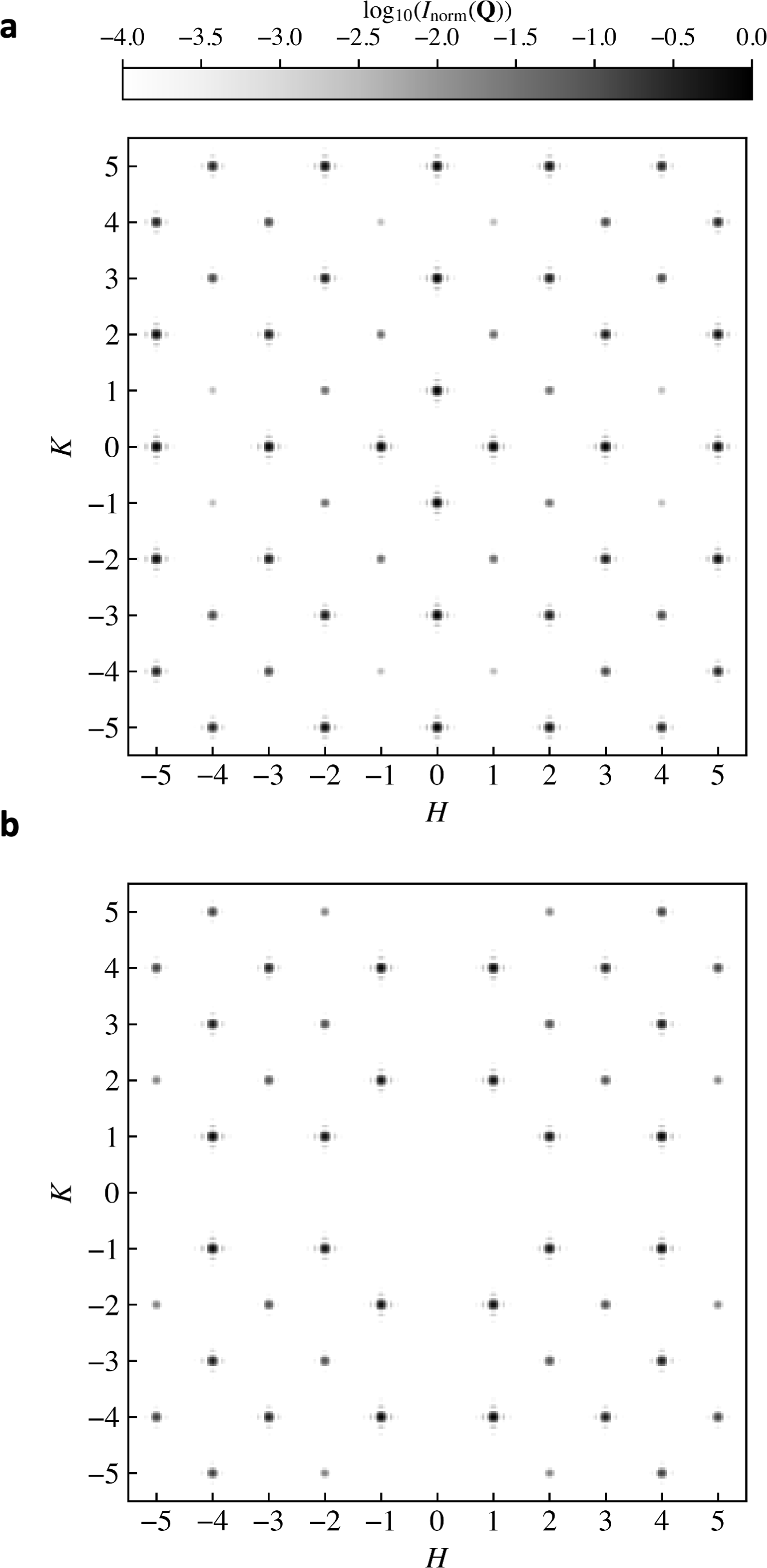}
    \caption{\linespread{1.0} \footnotesize {\bf SSFs for parallel and perpendicular spin components.} Simulated SSF maps for an infinite Type-I vertex configuration in a square ASI lattice obtained by considering, {\bf a}: the spin component projected parallel and {\bf b}: the perpendicular to $\mathbf{Q}$ according to Eq.~\eqref{SSF_par}.}
    \label{Appendix_extra_SSF}
\end{figure}

\noindent Here $N$ is the number of islands and $\mathbf{r}$ their real-space position vectors. Historically, neutron scattering has been extensively employed to probe spin-ice systems, with the measured magnetic structure factor determined by the component of the moment \emph{perpendicular} to the scattering vector $\mathbf{Q}$, i.e., $S^{\perp}$ \cite{Bramwell_Science_2001}. In contrast, resonant X-ray scattering is sensitive to the component of the magnetization \emph{parallel} to $\mathbf{Q}$. To assess how this difference in probe sensitivity affects the observed SSF, we calculate Eq.~\eqref{SSF_par} using either the parallel or the perpendicular spin components and compare the resulting maps in Supplementary Fig.~\ref{Appendix_extra_SSF}. 

\noindent
 
Systematic absences and/or weak intensities are observed at specific reciprocal lattice points in the simulated SSF maps (Supplementary Fig.~\ref{Appendix_extra_SSF}). In the case where only the spin components parallel to $\mathbf{Q}$ are considered, the reduced symmetry of the magnetic periodicity changes the selection rules that govern the appearance (or suppression) of the magnetic peaks, as described above. This symmetry dependence highlights the importance of accounting for probe-specific sensitivity when interpreting magnetic scattering data. In particular, understanding these symmetry constraints is essential both for optimizing experimental geometries and for accurately analyzing the intensity distributions in X-ray magnetic scattering patterns.

\subsection{Fe $L_{2,3}$ x-ray absorption}

The x-ray specular reflection of the continuous Fe-implanted sample was measured at the Sextants beamline at the Soleil synchrotron with an incidence angle of 5°. The results are presented in Supplementary Fig.~\ref{SI_XAS_5deg}. The iron $L_3$ absorption edge is clearly observed at 706.5 eV, while the $L_2$ edge appears at 721 eV.

\begin{figure}[h]
\includegraphics[width=7cm]{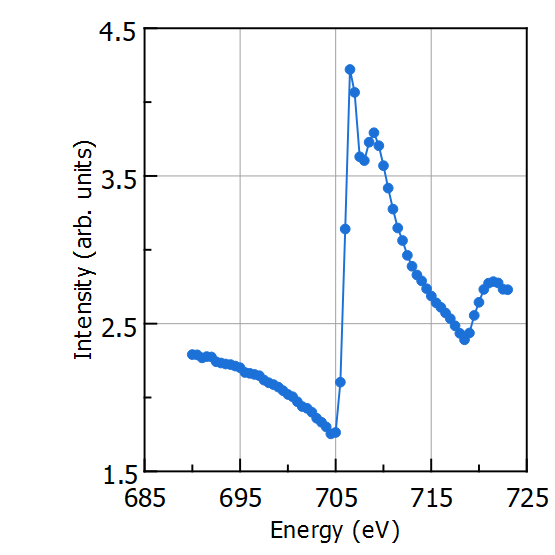}
\caption{\linespread{1.0} \footnotesize {\bf Intensity of the specular x-ray reflection at an angle of 5$^\circ$ for the Fe-implanted continuous film with nominal 40~nm Pd.} }
\label{SI_XAS_5deg}
\end{figure}

\end{document}